\documentclass[twocolumn, times]{aastex63}

\usepackage{amsmath,verbatim, nicefrac,color, soul}
\usepackage{natbib,graphicx}

\usepackage[caption=false]{subfig}
\newcommand\Hl[1]{\colorbox{yellow}}

\DeclareCaptionFormat{cont}{#1 (cont.)#2#3\par}

\shorttitle{Pulsar Cyclic Spectroscopy in the Partial--Deconvolution Regime}
\shortauthors{J. E. Turner \lowercase{et al}.}

\begin{document}

\title{Pulsar Cyclic Spectroscopy in the Partial--Deconvolution Regime: Benefits \& Limitations} 

\author[0000-0002-2451-7288]{Jacob E. Turner}
\affiliation{Green Bank Observatory, P.O. Box 2, Green Bank, WV 24944, USA}

\author[0000-0001-8885-6388]{Timothy Dolch}
\affiliation{Department of Physics, Hillsdale College, 33 E. College Street, Hillsdale, MI 49242, USA}
\affiliation{Eureka Scientific, 2452 Delmer Street, Suite 100, Oakland, CA 94602-3017, USA}

\author[0000-0002-6664-965X]{Paul B. Demorest}
\affiliation{National Radio Astronomy Observatory, 1003 Lopezville Rd., Socorro, NM 87801, USA}

\author[0000-0001-5229-7430]{Ryan S. Lynch}
\affiliation{Green Bank Observatory, P.O. Box 2, Green Bank, WV 24944, USA}

\author[0000-0002-1797-3277]{Daniel R. Stinebring}
\affiliation{Department of Physics and Astronomy, Oberlin College, Oberlin, OH 44074, USA}







\author[0000-0002-4188-6827]{Cody Jessup}
\affiliation{Department of Physics, Hillsdale College, 33 E. College Street, Hillsdale, MI 49242, USA}
\affiliation{Department of Physics, Montana State University, Bozeman, MT 59717, USA}

\author[0009-0009-7809-3335]{Nathaniel Jones}
\affiliation{Department of Physics, Hillsdale College, 33 E. College Street, Hillsdale, MI 49242, USA}

\author[0009-0007-2434-2776]{Christopher Scheithauer}
\affiliation{Department of Physics, Hillsdale College, 33 E. College Street, Hillsdale, MI 49242, USA}

\begin{abstract}
We explore possible advantages of cyclic spectroscopy for observations of pulsars in instances where full cyclic deconvolution is not possible. We compute cyclic merits and full-deconvolution regime boundaries for pulsars observed by NANOGrav and discuss which sources stand to benefit the most from using cyclic spectroscopy when observed with the Green Bank Telescope and DSA-2000 in a given frequency range. We compare data products, namely the wavefield, in both full-deconvolution and partial-deconvolution regimes to demonstrate what can be accomplished with incomplete phase retrieval. Additionally, we show how some phase retrieval can still be achieved in the partial-deconvolution regime and how this allows for additional information in scintillation-based data products, like the dynamic wavefield power, compared to what can be found in traditional dynamic spectra. An examination of dynamic wavefield phase as a function of observing frequency reveals more complete phase retrieval is achieved the closer one gets to the full deconvolution regime, agreeing with the expectations of cyclic merit. While we demonstrate that fragmentary recovery of the secondary wavefield can be accomplished in the partial-deconvolution regime, we advocate for a synergistic approach with phase retrieval methods like the $\theta-\theta$ transform, although we also provide discussion about shortcomings of this strategy. Finally, we use the combination of modest cyclic merit and lack of discernible results for PSR J1903$+$0327 to motivate the creation of an updated ``cyclic merit 2.0", which relies on scintillation bandwidth instead of observing bandwidth.
\end{abstract}

\section{Introduction}
\label{intro}
Cyclic spectroscopy is a data processing tool that has the potential to significantly impact how pulsar astronomy is performed in the near future. By taking advantage of the unique, naturally-occurring periodicity inherent in pulsar emission as viewed from Earth and resampling the data at harmonics of the inverse pulse period, one can bypass the time-frequency sampling uncertainty relation, also known as the Gabor limit, to achieve simultaneous high time (and thus, pulse phase) and frequency resolution in a single data product \citep{cyc_spec}. Additionally, unlike conventional spectroscopy, which can only recover the amplitude information inherent in an observed signal, cyclic spectroscopy can utilize the periodicity of pulsar emission to recover both amplitude and phase information -- in the case of pulsar astronomy, the amplitude and phase of the transfer function of the ionized interstellar medium (ISM) --  allowing for complete reconstruction of signals present in the data.
\par For a general cyclostationary electric field signal, $E(t)$, with periodicity, $P$, and frequency domain representation, $E(\nu)$, the namesake cyclic spectrum represents the information present in the signal at all observing frequencies, $\nu$, and harmonics, also referred to as cyclic frequencies, $\alpha_k$, where $\alpha_k=k/P$ and $k$ is an integer, and is given by 
\begin{equation}
    S_{E}(\nu,\alpha_k) = \langle E(\nu+\alpha_k/2)E^*(\nu-\alpha_k/2)\rangle.
\end{equation}
Here, bracket angles indicate expectation value and stars indicate complex conjugates. In the case of pulsar emission propagating through the ionized interstellar medium, the resulting cyclic spectrum can be written as
\begin{equation}
    S_{E}(\nu,\alpha_k) = \langle H(\nu+\alpha_k/2)H^*(\nu-\alpha_k/2)\rangle S_x(\nu, \alpha_k),
\label{pulse_cs}
\end{equation}
where $S_x(\nu, \alpha)$ is the Fourier transform of the intrinsic pulse profile and $H(\nu)$ is the transfer function, or frequency response, of the interstellar medium, which is imparted onto the intrinsic pulse during the latter's propagation through the medium \citep{wdv13}. In the time domain, the analogous pulse broadening function, $h(t)$, is seen as a convolution with the intrinsic pulse and is often visible in the resulting signal as a scattering tail. Recovery of this function allows for a direct measurement of scattering-based delays in pulse arrival time, and can significantly increase timing precision in many pulsars, improving pulsar timing array sensitivity to gravitational waves \citep{turner_cyc}. In real data, Equation \ref{pulse_cs} is only valid up to timescales dependent on the decorrelation of diffractive scintillation in the observed signal, and therefore must be resampled at least as often as every such scintillation timescale, $\Delta t_{\rm d}$.
\par The transfer function is of particular importance both to studies of the interstellar medium and to efforts by pulsar timing arrays to detect nanohertz frequency gravitational waves; it contains information allowing for study of the structure of the interstellar medium along pulsar lines of sight, as well as means to quantify the pulse time-of-arrival delays, $\tau$, caused by the emission passing through an ionized medium. When measured across many scintillation timescales, one can reconstruct the dynamic wavefield, $H(\nu,t)$, which can also be thought of as the time-evolving transfer function. This complex function is a time and frequency-evolving interference pattern that forms for an observer due to the multipath propagation of pulsar emission through an ionized medium, and is invaluable for probing the interstellar medium on astronomical unit-to-parsec scales, and is the preferred data product for pulsar scintillometry \citep{baker_2021}. 
\par Because this signal is difficult to directly recover and separate from the intrinsic pulse in standard Fourier spectroscopy-processed observations, efforts to achieve the aforementioned benefits are often limited to techniques that use approximations of the transfer function's intensity, which only utilizes amplitude information. The most common such data product is the pulsar dynamic spectrum, which tracks changes in the observed signal intensity over observing frequency and time. The fringe features in this interference pattern, known as scintles, can also be used to probe the structure of the interstellar medium, and their characteristic frequency widths, $\Delta \nu_{\rm d}$, also known as scintillation bandwidths, can be used to approximate pulse time-of-arrival delays due to interstellar scattering. 
\par By taking the squared modulus of the two-dimensional Fourier transform of the dynamic spectrum, one can create the secondary spectrum, which often has parabolic features known as scintillation arcs \citep{OG_arcs}. These structures are indicative of highly localized structures along the pulsar emission propagation path that are responsible for the dominant portion of scattering, and are comprised of a series of  images formed from interference between pulsar emission propagating along the line of sight and emission being scattered at a distribution of angles upon passing through the phase screen. This spectrum serves as the amplitude-derived analog of the secondary wavefield, which can be found in a similar manner by starting with the dynamic wavefield \citep{baker_2023}. This data product constrains scattered images to singular points in the spectrum, allowing for highly precise tracking of structures in the interstellar medium over time as they transverse our line of sight.
\par A multitude of approaches have been developed to facilitate transfer function and dynamic wavefield recovery, ranging from implementations of the CLEAN algorithm \citep{walker_stinebring, Young_2024}, to interstellar holography \citep{holography}, to utilization of giant pulses \citep{Mahajan_2023}. One of the more popular approaches in pulsar scintillometry is recreating the dynamic wavefield from dynamic and secondary spectra via what is known as the $\theta-\theta$ transform \citep{theta_theta}. With this technique, all points in the secondary spectrum are mapped to a space showing the distribution of scattered images formed from the interference between photons reaching our detectors from angles $\theta_1$ and $\theta_2$ under the assumption of 1D scattering. The dynamic and secondary wavefields can then be reconstructed by inversely mapping the images from the $\theta-\theta$ spectrum \citep{baker_2021}.
\par Unlike some of these methods, which can rely on data products that do not contain any intrinsic phase information, by processing data with cyclic spectroscopy, one can directly recover the transfer function from the cyclic spectrum. However, a potential limitation with this technique is the regime in which complete deconvolution of the transfer function can be accomplished. It has been demonstrated that cyclic spectra are only sensitive to phase differences on frequency scales up to the inverse of the intrinsic pulse width \citep{dsj+20}. In other words, complete deconvolution with this technique is only possible if scintles are smaller than this limit, or, equivalently, one can see a scattering tail in the pulse profile, assuming sufficient signal-to-noise (S/N). However, in this paper, we make the argument that, assuming one can process their data with cyclic spectroscopy, there are still a multitude of significant benefits to be gained by using this technique over conventional Fourier spectroscopy in data that falls outside of the full-deconvolution regime.
\par In Section \ref{sec:data} we present data used for these analyses. In Section \ref{sec:analyses} we discuss our analyses, including determination of deconvolution regime and cyclic merit, data processing, and construction of wavefields and spectra. In Section \ref{sec:results}, we present the results of our calculations and compare these various data products. In Section \ref{sec:discussion}, we provide detailed discussion of these results, implications for the future of pulsar astronomy, and general strategies for technique adoption and pulsar selection. Finally, in Section \ref{concl}, we present our conclusions and discuss the future of the field, with an emphasis on the forthcoming availability of a cyclic spectroscopy backend. 
\section{Data}
\label{sec:data}
To best facilitate discussion of this cyclic spectroscopy use scenario, our paper utilizes data from three separate raw voltage, i.e., baseband, campaigns spanning almost 20 years in time and almost 1500 MHz in observing bandwidth. Our earliest data are an observation of the millisecond pulsar PSR B1937$+$21 (P2067 PI Demorest \citep{cyc_spec,wdv13}) taken with the ASP spectrometer on the Arecibo Telescope. Our second set of observations were also of PSR B1937$+$21 (P2676 PI Dolch, \citep{dsj+20, Turner_1937}) and were taken with the PUPPI spectrometer on the Arecibo Telescope. Our final set of observations were of the millisecond pulsars PSRs J1643--1224, J1903$+$0327, and B1937$+$21 (GBT24B$\underline{\hspace{0.2cm}}$039 PI Turner) taken with the VEGAS spectrometer on the Green Bank Telescope (GBT) across 9 epochs, with each epoch dedicated to a single pulsar. All observations were taken in the ``raw" pulsar observing mode, as detailed in the observing guides for PUPPI\footnote{\url{https://www.cv.nrao.edu/~pdemores/puppi/}} and VEGAS\footnote{\url{https://www.gb.nrao.edu/scienceDocs/GBTog.pdf}}.
\par As discussed in Section \ref{sec:discussion}, RFI mitigation is generally not required for phase-recovered cyclic spectroscopy data products. However, for analyses involving amplitude-only data products such as dynamic spectra and intensity pulse profiles, we excised RFI using median smoothed difference zapping via the ``paz --r" command in \texttt{psrchive}\footnote{\url{https://psrchive.sourceforge.net/}} after the data had been processed with cyclic spectroscopy. Since none of our observations were taken with overlapping polyphase filterbank (PFB) channels, gaps between channels resulted in a loss of signal in stretches of frequency space approximately 5\% of the channel width. To mitigate this issue, PFB channel widths were chosen such that they were significantly larger than predicted scintillation bandwidths, resulting in a minimal chance of scintillation structures being intersected. Information on the observing setups and dates can be found in Table \ref{scint_table}. 

\begin{deluxetable*}{CCCCCCCCC}

\tablewidth{\columnwidth}
\tablecolumns{7}

\tablecaption{Observation Setups \label{scint_table}}
\tablehead{ \colhead{Pulsar} & \colhead{Day(s) Observed} & \colhead{Center Freq.} & \colhead{Obs. BW} & \colhead{PFB BW} & \colhead{Cyclic Resolution} &  \colhead{Subint. Length} & \colhead{Obs. Length} \\ \colhead{} & \colhead{(MJD)} & \colhead{\text{(MHz)}} & \colhead{\text{(MHz)}} & \colhead{(MHz)} & \colhead{\text{(kHz)}} & \colhead{\text{(sec)}}  & \colhead{\text{(min)}}  \vspace{0.05cm}}
\startdata
\rm{B}1937{+}21$^a$ &   53873 & 428 & 4 & 4 & 0.98 & 15 & 60 \\
\rm{B}1937{+}21$^a$ &   56183, 56198, 56206 & 1373.125 & 200 & 6.25 & 6.1 & 28 & 120{$-$}150 \\  
\rm{B}1937{+}21$^g$ &   60581, 60693 & 1500 & 800 & 25 & 6.1 & 25 & 90{$-$}120 \\  
\rm{J}1643{$-$}1224$^g$ & 60540, 60602, 60635,60689 & 1500 & 800 & 1.5625 \text{ or } 25 & 6.1 & 25 & 90{$-$}120 \\  
\rm{J}1903{+}0327$^g$ & 60539, 60577, 60607 & 1500 & 800 & 1.5625  & 6.1 & 25 & 90{$-$}120
\enddata
\tablecomments{Observing information for all pulsars used in this study.\\$^{\rm a}$Observed with Arecibo. \\ $^{\rm g}$Observed with GBT.}
\end{deluxetable*}

\section{Analyses}
\label{sec:analyses}
\subsection{Initial Cyclic Spectral Processing}
All data were processed via cyclic spectroscopy using \texttt{dspsr}\footnote{\url{http://dspsr.sourceforge.net/}} \citep{dspsr} with 1024 pulse phase bins. Details regarding data resolution post-cyclic processing can be found in Table \ref{scint_table}. Of particular note is that all L--band data has 6.1 kHz frequency resolution, which is a factor of 256 better than what is used in NANOGrav's current observing setup in that frequency range, while simultaneously maintaining comparable pulse phase resolution \citep{Agazie_2023_timing}.

\subsection{Cyclic Phase Retrieval and Wavefield Reconstruction}
Multiple approaches have been developed to recover the transfer function from the cyclic spectrum. \cite{turner_cyc} demonstrated that one can reconstruct the transfer function assuming the cyclic spectrum and transfer function amplitudes are approximately equivalent and that the transfer function phase can be recovered by integrating over the cyclic spectrum phase with respect to frequency. In this work, we utilize the iterative deconvolution approach available in the python package \texttt{pycyc}
\footnote{\url{https://github.com/gitj/pycyc}}. The process for then reconstructing our data's wavefields was as follows: For each subintegration in a given observation, we used \textsc{pycyc} to recover the transfer function. We then stacked these transfer functions in time to create the dynamic wavefield for that observation. From there, we created the dynamic wavefield power by taking the squared modulus of the dynamic wavefield. Finally, the secondary wavefield for a given observation was created by taking the squared modulus of the 2D Fourier transform of the dynamic wavefield.

\subsection{Dynamic and Secondary Spectra Creation}
In addition to phase-recovered signals, we also constructed our data's amplitude-only signals from the cyclic spectroscopy-processed filterbank data. To create each epoch's dynamic spectrum, after summing our data over all polarizations, we created an average profile of the observation and subsequently integrated over the product of this profile and the observed pulse profile at each instance in subintegration-frequency space, resulting in the measured summed power over the on-pulse window. We then subtracted the off-pulse window, which constituted the remainder of the pulse profile, and divided the resulting signal by the baseline intensity of the pulse profile. At higher observing frequencies in some observations of PSR J1643--1224, our S/N was too low to visualize scintles with our full pulse phase and frequency resolution, and so we averaged to 32 pulse phase bins prior to creating dynamic spectra in these instances in order to improve S/N when integrating over the pulse.

\par To create secondary spectra, common practice involves taking the squared modulus of the 2D Fourier transforms of their dynamic spectra. However, given that both dynamic spectra and dynamic wavefield power are amplitude-only functions, in cases where S/N was sufficient for attempted deconvolution, we instead opted to create our secondary spectra from the squared modulus of the 2D Fourier transforms of our dynamic wavefield power arrays. Otherwise, we used the former approach.

\subsection{Estimation of Deconvolution Regime \& Cyclic Merit}
\subsubsection{Our Observations}
To determine whether we should expect phase retrieval for each set of observations, we first approximated its corresponding deconvolution regime based on whether that data satisfied the relation
\begin{equation}
    \Delta \nu_{\rm d} \lesssim \frac{1}{W_{10}},
\end{equation}
where $W_{10}$ is the width of the pulse at 10\% of the maximum. This choice of cutoff reflects cyclic spectra only being sensitive to phase differences on frequency scales up to $1/W$ as mentioned earlier, where $W$ is the intrinsic pulse width.
\par Our scintillation bandwidths and timescales were determined by taking subbands of processed dynamic spectra (or dynamic wavefields, when available) over which we expected minimal change in scintle size and obtaining their 2D autocorrelation functions (ACFs) using \textsc{pypulse} \citep{pypulse}. We then took the 1D slice at zero frequency and time lag to acquire the 1D frequency and time ACFs, respectively. Finally, we fit the frequency ACFs with Lorentzians and found the scintillation bandwidths by taking the half width at half maxima of the resulting fits, and fit the time ACFs with one-sided exponentials and found the scintillation timescales by taking the half width at $e^{-1}$ of the resulting fits \citep{Cordes_1985}. The uncertainties in these measurements were given by a summation in quadrature of the fit uncertainty and the uncertainty from the finite number of scintles in the observation, the latter of which is determined via
\begin{equation}
\label{finite_scintle}
\begin{split}
\epsilon &  = \frac{\Delta \nu_{\rm d}}{2\ln(2) N_{\rm scint}^{1/2}} \\
& \approx \frac{\Delta \nu_{\rm{d}}}{2\ln(2)[(1+\eta_{\text{t}}T/\Delta t_\text{d})(1+\eta_\nu B/\Delta \nu_\text{d})]^{1/2}},
\end{split}
\end{equation}
where the total number of scintles is given by ${N}_{{\rm{scint}}}$, the total integration time by $T$, the observing bandwidth by $B$, and filling factors, typically between $0.1$ and $0.3$, and, this case, set to 0.2, are given by ${\eta }_{{\rm{\nu}}}$ and ${\eta}_{t}$ \citep{Cordes1986}.
\par To determine the likelihood of achieving successful phase retrieval, we calculated the cyclic figure of merit, $m_{\rm cyc}$ \citep{dsj+20}, for each observation in which we calculated a scintillation bandwidth. This metric is given by
\begin{equation}
\label{cyc_merit}
m_{\rm cyc} = \frac{\Phi}{\delta \Phi}=2\pi\frac{\tau_{\rm d}W_e}{P^2}(\textrm{S/N}) \sqrt{\sum_k k^2 a_k},
\end{equation}
where $W_e$ is the equivalent pulse width, $\Phi$ is the cyclic phase, $\tau_{\rm d}$ is the scattering delay, and $a_k \equiv A_k/A_0$, where $A_k$ is the $k^{\rm th}$ amplitude of the intensity pulse profile's Fourier transform. We chose to approximate $W_e$ with the effective pulse width, $W_{\rm eff}$, although using $W_{10}$ yielded very similar cyclic merits for these sources. In general, one should expect successful deconvolution for $m_{\rm cyc} \gg 1$. For a given observation, we determined this quantity only over the subintegration times mentioned in Table \ref{scint_table}. 
\par For our study, we approximated scattering delay using the relation
\begin{equation}
\label{uncertainty_scat}
    \Delta \nu_{\rm d} = \frac{C_1}{2\pi\tau_{\rm d}},    
\end{equation}
where $C_1$ can typically range from around $0.6-1.5$ depending on factors including the geometry of the corresponding scattering screen, as well as the assumed turbulence of the line of sight's electron density wavenumber spectrum \citep{Cordes_1998}. Assuming thin screen scattering with Kolmogorov turbulence, we get $C_1=0.957$. For a given observation and frequency range in which a scintillation bandwidth measurement was made, we initially calculated S/N by taking the ratio of the on-pulse maximum and the off-pulse RMS individually in all subintegrations. We then used a S/N and uncertainty equal to the mean and standard deviation, respectively, of the resulting measurements.  Finally, we used the approximation of Equation \ref{cyc_merit}'s radical from \cite{turner_cyc}, where
\begin{equation}
    \label{approx}
    \begin{split}
    \sqrt{\sum_k k^2 a_k} & \approx \sqrt{\sum_{k=1}^{k_{\textrm{max}}} k^2} = \sqrt{\frac{k_{\textrm{max}}(k_{\textrm{max}}+1)(2k_{\textrm{max}}+1)}{6}} \\ & \approx \sqrt{k_{\textrm{max}}^3}\approx \Bigg(\frac{P}{W_e}\Bigg)^{3/2}.
    \end{split}
\end{equation}
\par We note that Equation \ref{cyc_merit} is continuous, and increasing factors like scattering delay or S/N will improve one's chances at a successful cyclic deconvolution, and, as demonstrated in this paper, result in gradually better phase retrieval. However, we still believe that discussion and framing in terms of full and partial deconvolution is valuable, as in the vast majority of cases it will likely be clear cut whether an observation has achieved full (or effectively full) phase retrieval. In the event of borderline cases, for the purposes of discussing the benefits of the different regimes, we feel it makes more sense to treat them in the context of full deconvolution, especially if they have a high cyclic merit. It is likely that, in real data, even in the full regime, there will almost certainly be some small degree of incomplete phase retrieval, even if just in a trivial sense.
\subsubsection{NANOGrav Observations}
\par To explore how cyclic merit and the threshold to the full deconvolution regime apply to current pulsar timing array programs, we also determined these two quantities using the scintillation bandwidths and scattering delays measured in the NANOGrav 9- and 12.5-year data sets \citep{Levin_Scat, turner_scat}. For pulsars in \cite{turner_scat} or \cite{Levin_Scat} without measured scintillation bandwidths, we used predictions from the NE2001 electron density model \citep{NE2001}, except in the case of the three pulsars observed for this paper, where we simply used our measured values, and assumed a 50\% variation in scattering delays from epoch to epoch. We used $W_{10}$ values taken from \cite{stairs}, \cite{Jacoby_2007}, \cite{2013_width}, \cite{2011_pol}, \cite{2020_width}, \cite{fast}, \cite{2024_deneva}, and \cite{2024_wang} and $W_{e}$ values from \cite{Lam_2019}. 
\par Rather than going through all epochs to estimate S/N using the approach described above, we instead opted to estimate typical source S/N via
\begin{equation}
    \label{s_n_estimate}
    \textrm{S/N} = \frac{SG}{T}\sqrt{2\Delta t_{\rm d} \textrm{BW}}\sqrt{\frac{P-W_{10}}{W_{10}}},
\end{equation}
where $S$ is pulsar flux density, $G$ is telescope gain, $T$ is system temperature, and BW is the observing bandwidth over which we can reasonably expect minimal scintle evolution, which we assumed to be 5 MHz, 50 MHz, and 100 MHz at observing frequencies of 350 MHz, 820 MHz, and 1500 MHz, respectively. We used measured pulsar flux densities taken from \cite{alam2020nanograv}, with uncertainties given by the 16$^{\rm th}$ and 84$^{\rm th}$ percentile measurements. If we ran analyses in frequency ranges for which measured flux densities were not available, we used the provided spectral indices to estimate flux densities at these new observing frequencies. We also adopted the same policy regarding scintillation timescale measurements as we did for scintillation bandwidth and scattering delay measurements. For the various frequencies at which we performed our analyses, we scaled our scintillation parameters assuming $\nu^{1.2}$ frequency dependence for scintillation timescale and $\nu^{4.4}$ dependence for scintillation bandwidth. 
\par Given the imminent utilization of next-generation telescopes by pulsar timing arrays, we opted to determine cyclic merits both for GBT and for the upcoming DSA-2000, which is expected to have Arecibo levels of sensitivity. For GBT, we assumed all observing frequencies would have a gain of 2 K/Jy and system temperatures of 25, 30, and 18 K at 350, 820, and 1500 MHz, respectively. For DSA-2000, we assumed a gain of 10 K/Jy and a system temperature of 25 K. These system temperatures were added to the frequency-dependent temperature contributions from the galactic background at each pulsar's location on the sky, as provided by \textsc{PyGDSM} \citep{2008MNRAS.388..247D, 2015MNRAS.451.4311R, 2016ascl.soft03013P, 2017MNRAS.464.3486Z, 2017MNRAS.469.4537D}, to get the final temperature used in our calculations.
\section{Results}
\label{sec:results}
\subsection{Cyclic Regime \& Likelihood of Deconvolution Success}
\subsubsection{Our Observations}
\begin{deluxetable*}{CCCCCCCC}[!ht]
\centering

\tablecolumns{8}

\tablecaption{Cyclic Regime \& Merit \label{regime_table}}
\tablehead{\colhead{Pulsar} & \colhead{Center Freq.} & \colhead{Obs. BW} & \colhead{$\Delta \nu_{\textrm{d}}$} & \colhead{Regime} & \colhead{Max Obs. Freq. for Full Regime} & \colhead{$m_{\rm cyc}$} & \colhead{S/N for $m_{\rm cyc}=20$}
\\ \colhead{} & \colhead{\text{(MHz)}} & \colhead{\text{(MHz)}} & \colhead{\text{(kHz)}} & \colhead{} & \colhead{(MHz)} & \colhead{} & \colhead{}\vspace{0.05cm}}
\startdata
\rm{B}1937{+}21 \ (\rm{Arecibo}) & 428 & 4 & 7.4 $\pm$ 0.3$^w$ & \rm{Full} & 473 $\pm$ 4 & 15 $\pm$ 2$^*$ & 114 $\pm$ 5$^*$ \\
\rm{B}1937{+}21 \ (\rm{Arecibo}) & 1310 & 25 & 398 $\pm$ 23$^{w}$ & \rm{Partial} & 585 $\pm$ 8 & 0.24 $\pm$ 0.05$^*$ & 5400 $\pm$ 300$^*$ \\
\rm{B}1937{+}21 \ (\rm{Green \ Bank}) &  1324 & 22 & 461 $\pm$ 41$^w$ & \rm{Partial} & 572 $\pm$ 12 & 0.12 $\pm$ 0.02$^*$ & 6300 $\pm$ 600$^*$\\
\rm{J}1643{$-$}1224 \ (\rm{Green \ Bank}) & 1324 & 22  & 29.6 $\pm$ 0.3 & \rm{Partial} & 623 $\pm$ 1 & 0.09 $\pm$ 0.01 & 1300 $\pm$ 10\\
\rm{J}1903{+}0327 \ (\rm{Green \ Bank})  & 1250 & 33 & 0.86$\pm 0.01^a$ & \rm{Full} &  1360 $\pm$ 4 & 4.1 $\pm$ 0.4 & 16 $\pm$ 1 
\enddata
\tablecomments{A representative sample of deconvolution regime determinations and cyclic merit calculations from our various observations. Cyclic merit was determined over subintegration times shown in Table \ref{scint_table}. Maximum observing frequency for observation to be in the full deconvolution regime was determined by assuming scintillation bandwidth scales as $\nu^{4.4}$. $W_{\rm eff}$ and $W_{10}$ for PSR B1937+21 were taken from \cite{stairs}. $W_{\rm eff}$ and $W_{10}$ for PSR J1643--1224 were taken from \cite{Lam_2019} and \cite{2011_pol}, respectively. $W_{\rm eff}$ and $W_{10}$ for PSR J1903+0327 were taken from \cite{Lam_2019} and \cite{fast}, respectively. \\$^*$Stated cyclic merit for B1937+21 is half as large as $m_{\rm cyc}$ calculation due to the interpulse \citep{dsj+20}.\\$^a$Estimated from pulse broadening-inferred scattering delay taken from \cite{geiger2024nanograv125yeardataset} at the same observing frequency.\\$^w$Scintillation bandwidths measured using dynamic wavefield power.}
\end{deluxetable*}
We determined the cyclic figure of merit within the previously mentioned subintegration times for all pulsars in our data set, along with estimations of their full deconvolution regimes and required S/N to achieve a reasonably high cyclic merit of 20, which satisfies the requirement of $m_{\rm cyc}\gg 1$. That being said $m_{\rm cyc} \gtrsim 10$ is probably sufficient, and it has been suggested that in some cases deconvolution may be possible with $m_{\rm cyc} < 10$ \citep{dsj+20}, although this would likely require a source to be observed in the full deconvolution regime and achieve a moderate S/N. For example, our 428 MHz data for PSR B1937+21 is well within the full deconvolution regime, and, for the same scintillation bandwidth we measured, would have a cyclic merit of around 6 if we lowered its S/N to 30, which is still likely sufficient for successful deconvolution. Some examples of these measurements are shown in Table \ref{regime_table}.
\par We chose to measure cyclic merits only within individual subintegration times, as this allows for the determination of deconvolution likelihood while maintaining the temporal resolution required to sufficiently reconstruct a given observation's dynamic wavefield. However, if all we desire is a reconstructed pulse broadening function, then we can improve our S/N, and, consequentially, cyclic merit, by integrating over lengths of data equivalent to an observation's scintillation timescale. These cyclic merits are shown in Table \ref{timescale_table} for the corresponding observations from Table \ref{regime_table}. While these longer integration times noticeably improve the cyclic merit for every pulsar for which the scintillation timescale was already longer than its subintegration length, it is not enough for one to expect successful deconvolution relative to a single subintegration for these pulsars in instances where cyclic deconvolution would not have been previously achieved. For some pulsars with longer scintillation timescales relative to reasonable subintegration lengths, these longer integration times may significantly improve cyclic merit. However, this is unlikely to impact most of the pulsars in frequency regimes for which we expect cyclic deconvolution, as these pulsars will be highly scattered, and scattering delay is expected to be negatively correlated with scintillation timescale \citep{Cordes_1998}.

\begin{deluxetable}{CCC}[!ht]
\centering

\tablecolumns{3}
\tablecaption{Cyclic Merit Over Scintillation Timescales \label{timescale_table}}
\tablehead{\colhead{Pulsar} & \colhead{Center Freq.} & \colhead{$m_{\rm{cyc},\Delta t_{\rm d}}$} 
\\ \colhead{} & \colhead{\text{(MHz)}} & \colhead{} \vspace{0.05cm}}
\startdata
\rm{B}1937{+}21 \ (\rm{Arecibo}) & 428 & 29 $\pm$ 3$^*$\\
\rm{B}1937{+}21 \ (\rm{Arecibo}) & 1310 & 0.58 $\pm$ 0.09$^*$\\
\rm{B}1937{+}21 \ (\rm{Green \ Bank}) &  1324 & 0.27 $\pm$ 0.05$^*$\\
\rm{J}1643{$-$}1224 \ (\rm{Green \ Bank}) & 1324 & 0.12 $\pm$ 0.01 \\
\rm{J}1903{+}0327 \ (\rm{Green \ Bank})  & 1250 & <4.1 $\pm$ 0.4$^{\dagger,n}$  
\enddata
\tablecomments{Cyclic merits for the pulsars shown in Table \ref{regime_table} over full scintillation timescales rather than individual subintegrations.\\$^*$Stated cyclic merit for B1937+21 is half as large as $m_{\rm cyc}$ calculation due to the interpulse \citep{dsj+20}.\\$^{\dagger}$ Subintegration times are longer than scintillation timescale. \\$^n$ Scintillation timescale estimated from NE2001 electron density model.}
\end{deluxetable}
\par While two of our sets of observations (PSR B1937+21 at 428 MHz and PSR J1903+0327 at L-band) were in the full deconvolution regime for their respective pulsars, only one set had sufficient S/N and $m_{\rm cyc}$ to achieve successful deconvolution. We can also see for PSR B1937+21 that the approximate observing frequency threshold for the full deconvolution regime varies over time along with fluctuations in the scintillation bandwidth, which will be important to keep in mind when planning observations with the intent of achieving full cyclic deconvolution in the resulting data. Despite a large scattering delay, PSR J1643--1224's low S/N would require instruments significantly more sensitive than the GBT for modest phase retrieval to be possible in the partial deconvolution regime. Additionally, while our results suggest that PSR J1903+0327 would likely achieve successful deconvolution with a modest pulse S/N of around 16, the incredibly low S/N per scintle at the required frequency channelization to resolve these scintles would necessitate a more sensitive telescope than the GBT. This issue is discussed in more detail in Section \ref{sec:discussion}. While additional observations will be needed for confirmation, these findings indicate that an observation may be in the full deconvolution regime but may not have successful deconvolution due to a low cyclic merit, or, conversely, an observation may have a high cyclic merit but not achieve complete cyclic deconvolution because the observation was not taken in the full deconvolution regime. For this reason, both criteria should be kept in mind when evaluating or preparing to create cyclic spectroscopy-processed data.
\par
\begin{figure}[!ht]
    \centering
    \captionsetup[subfigure]{labelformat=empty}
    {\hspace*{-.8cm}\includegraphics[width=0.55\textwidth]{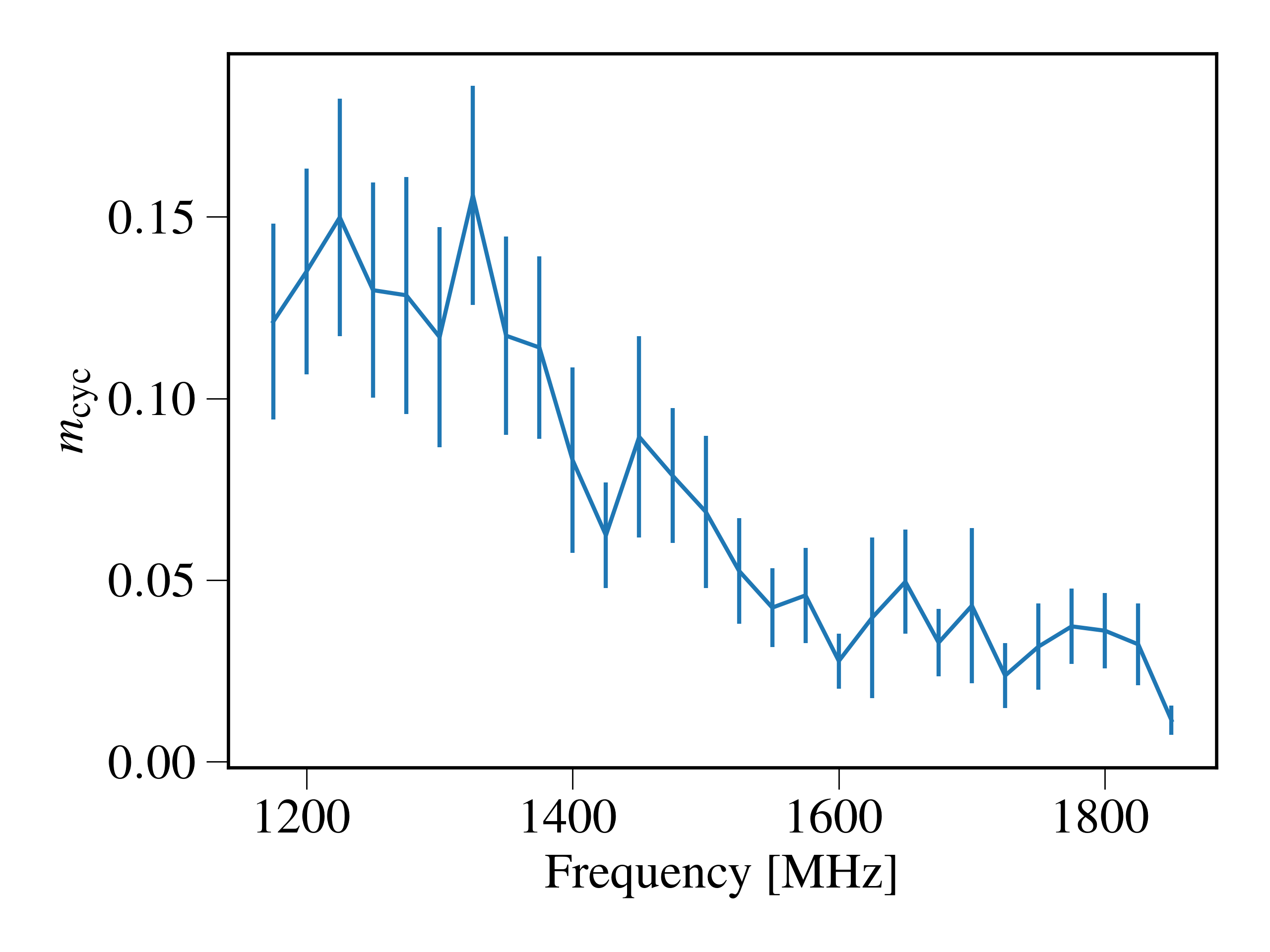} }%
    \caption{Cyclic merit over observing frequency for PSR B1937+21 on MJD 60581. This metric decreases as a source is observed further from the full-deconvolution regime.}%
    \label{merit_freq}%
\end{figure}

 As mentioned earlier, cyclic merit is a continuous function, and is expected to decrease as a source is observed further from the full-deconvolution regime. This results in lower cyclic merit at higher observing frequencies as scattering delay and pulsar flux density decline, as visualized in Figure \ref{merit_freq}.
\subsubsection{NANOGrav Observations}
\label{nano_cyc}
\par
\begin{figure*}[!htp]
    \centering
    \captionsetup[subfigure]{labelformat=empty}
    {\includegraphics[width=1.0\textwidth]{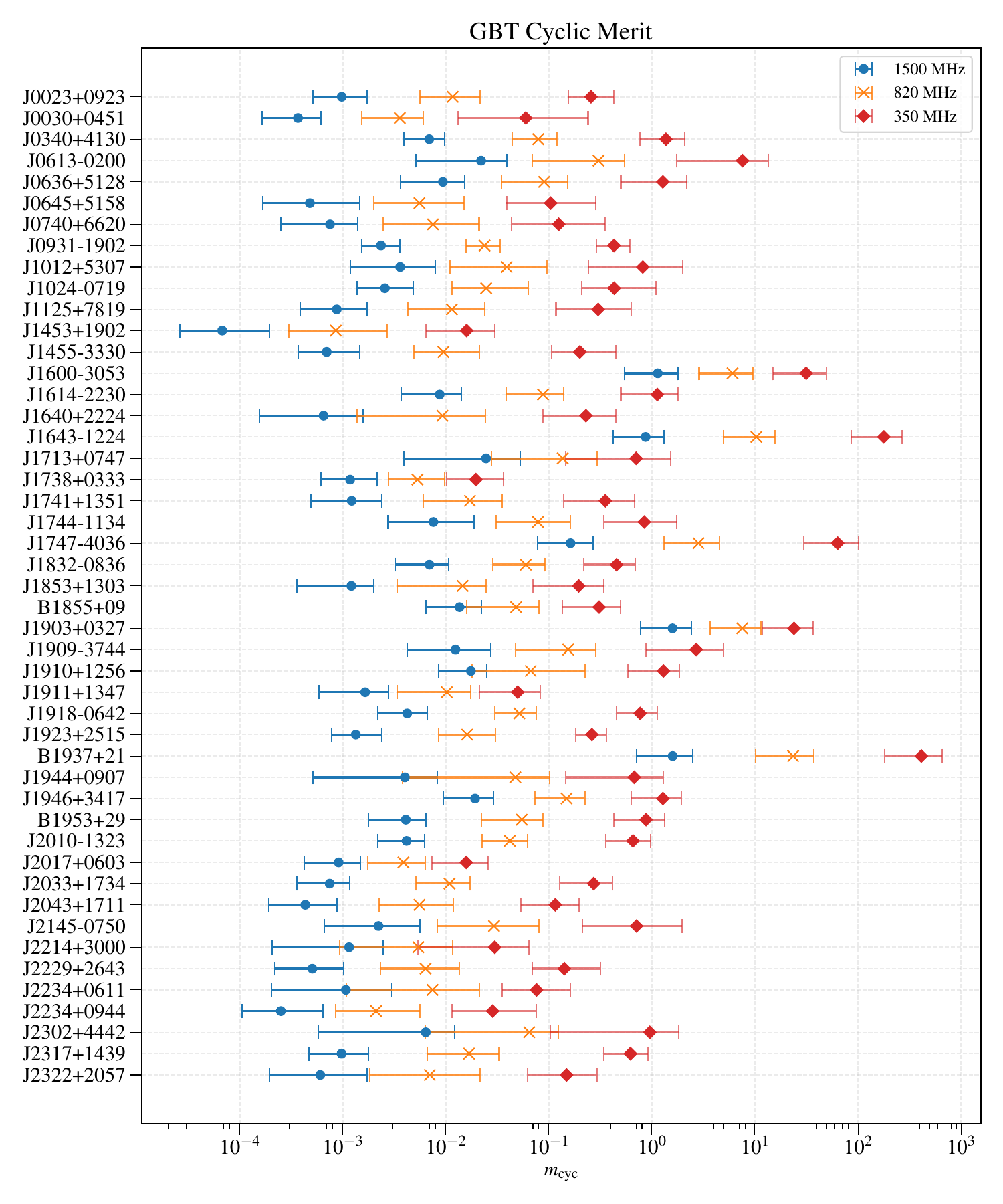} }%
    \caption{Estimated cyclic merits for pulsars in the NANOGrav 12.5-year data set at observing frequencies of 350 (red diamonds), 820 (orange crosses), and 1500 (blue circles) MHz assuming data were taken with the GBT. We would expect full cyclic deconvolution for sources with $m_{\rm cyc}\gg 1$.}%
    \label{gbt_merit}%
\end{figure*}

\begin{figure*}[!htp]
    \centering
    \captionsetup[subfigure]{labelformat=empty}
    {\includegraphics[width=1.0\textwidth]{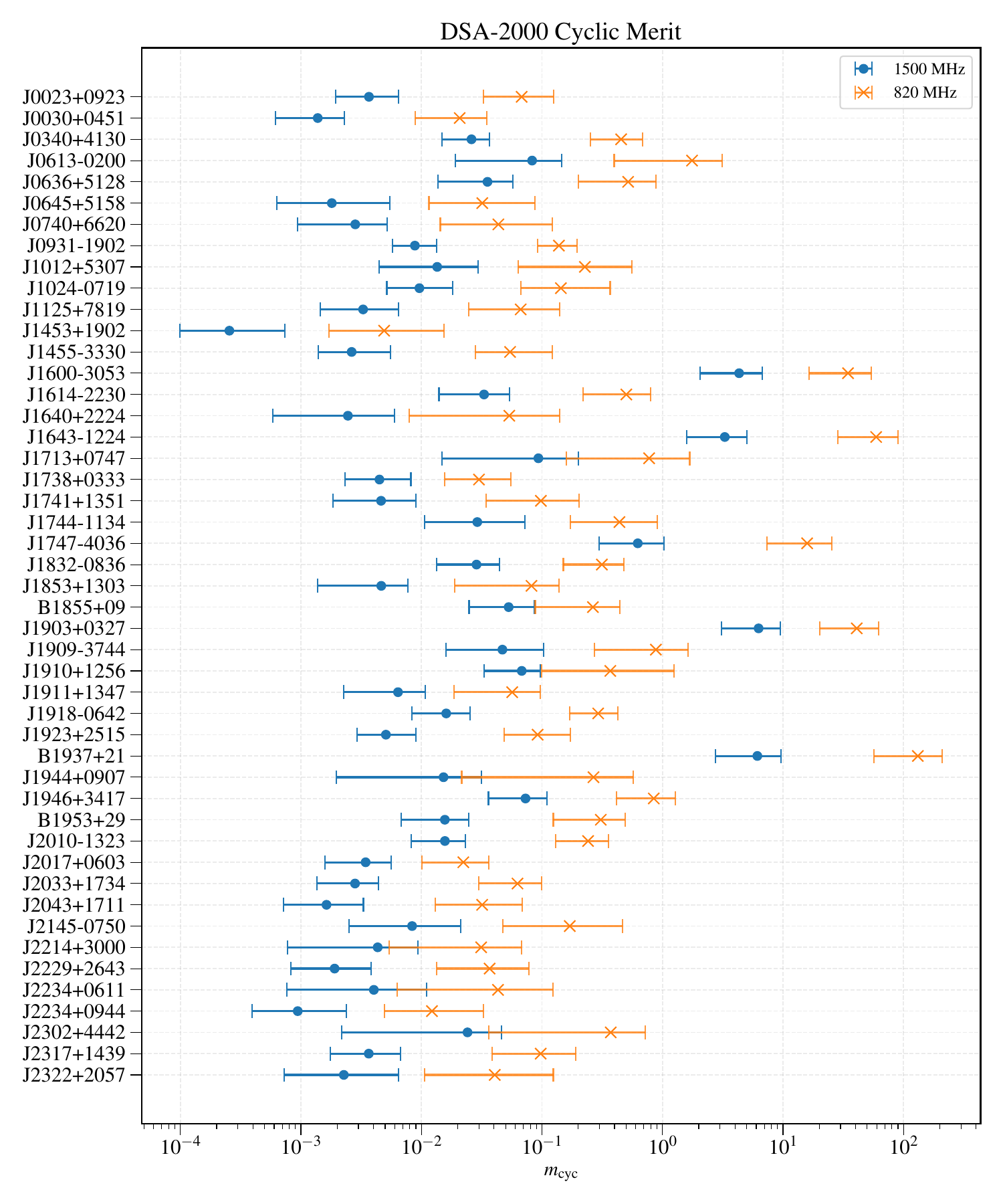} }%
    \caption{Estimated cyclic merits for pulsars in the NANOGrav 12.5-year data set at observing frequencies of 820 (orange crosses) and 1500 (blue circles) MHz assuming data were taken with DSA-2000. We would expect full cyclic deconvolution for sources with $m_{\rm cyc}\gg 1$.}%
    \label{dsa_merit}%
\end{figure*}

\begin{figure*}[!htp]
    \centering
    \captionsetup[subfigure]{labelformat=empty}
    {\includegraphics[width=1.0\textwidth]{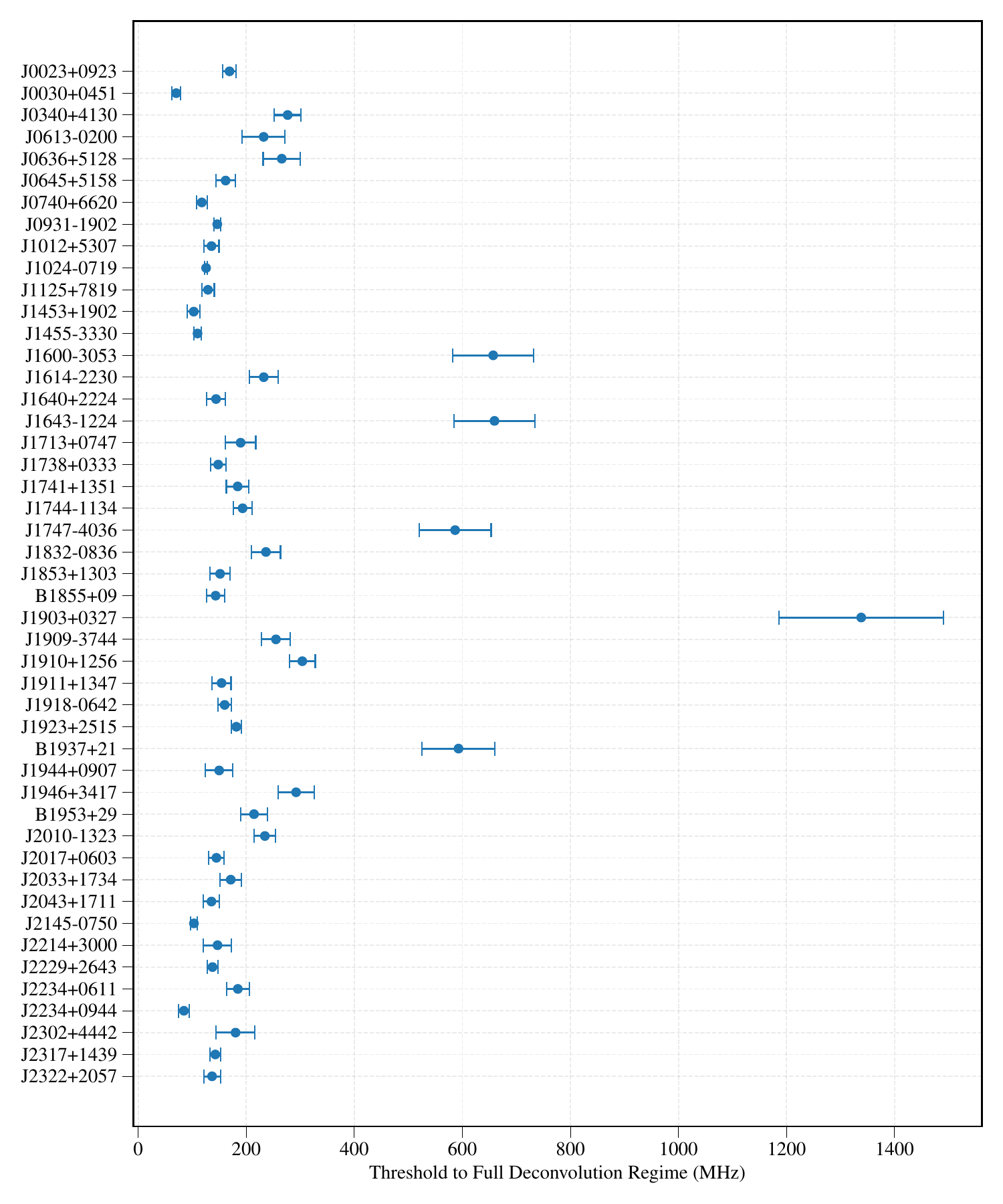} }%
    \caption{Estimated maximum observing frequency for pulsars in the NANOGrav 12.5-year data set to be observed in the full deconvolution regime.}%
    \label{decon_plot}%
\end{figure*}

 Cyclic merit estimations for NANOGrav pulsars when observed with the GBT or DSA-2000 can be seen in Figures \ref{gbt_merit} and \ref{dsa_merit}, respectively, while estimations of each pulsar's threshold to the full deconvolution regime are shown in Figure \ref{decon_plot}. These figures strongly suggest that the vast majority of pulsars that NANOGrav observed as part of its 12.5-year data release likely would not achieve successful cyclic deconvolution in any frequency range in which NANOGrav could reasonably observe, regardless of the telescope being used. However, six of these pulsars, PSRs J0613$-$0200, J1600$-$3053, J1643$-$1224, J1747$-$4036, J1903+0327, and B1937+21, pass the cyclic merit threshold of 10 at 350 MHz when observed with GBT, as well as at 820 MHz when observed with DSA-2000, with the exception of PSR J0613$-$0200. PSRs J1643$-$1224, J1903+0327, and B1937+21 also pass this threshold at 820 MHz when observed with the GBT. 
\par These results are not too surprising when one examines these cyclic merits alongside the corresponding full deconvolution regime boundaries, as Figure \ref{decon_plot} demonstrates that this regime is only accessible below 400 MHz for all but the pulsars mentioned above with the exception of PSR J0613$-$0200, who have regime boundaries between 400 and 1600 MHz, within error. These boundaries also suggest that, while significant phase retrieval would likely occur for these five pulsars if observed at 820 MHz with DSA-2000, full phase retrieval, and optimal reconstruction of the pulse broadening function, may not be possible at this observing frequency. 
\par While these results suggest that PTAs would need to observe in the full-deconvolution regime to achieve maximal benefit from this technique, these results in no way suggest that PTAs would not benefit from cyclic spectroscopy with their current observing strategies; any millisecond pulsar whose data were processed with this technique would still be able to achieve simultaneous high pulse phase and frequency resolution, which is crucial for estimating scattering delay in pulsars for which cyclic deconvolution is not possible at a given observing frequency, not to mention performing detailed studies of the ISM. Additionally, as discussed earlier in this subsection, and will be discussed further in the following subsection, cyclic deconvolution is a continuous process, and partial phase retrieval does occur outside of the full deconvolution regime, which still provides significant benefits.

\subsection{Dynamic Wavefields}
\begin{figure*}[!ht]
    \centering
    \captionsetup[subfigure]{labelformat=empty}
    \subfloat[]{ {\hspace{-1cm}\includegraphics[width=0.51\textwidth]{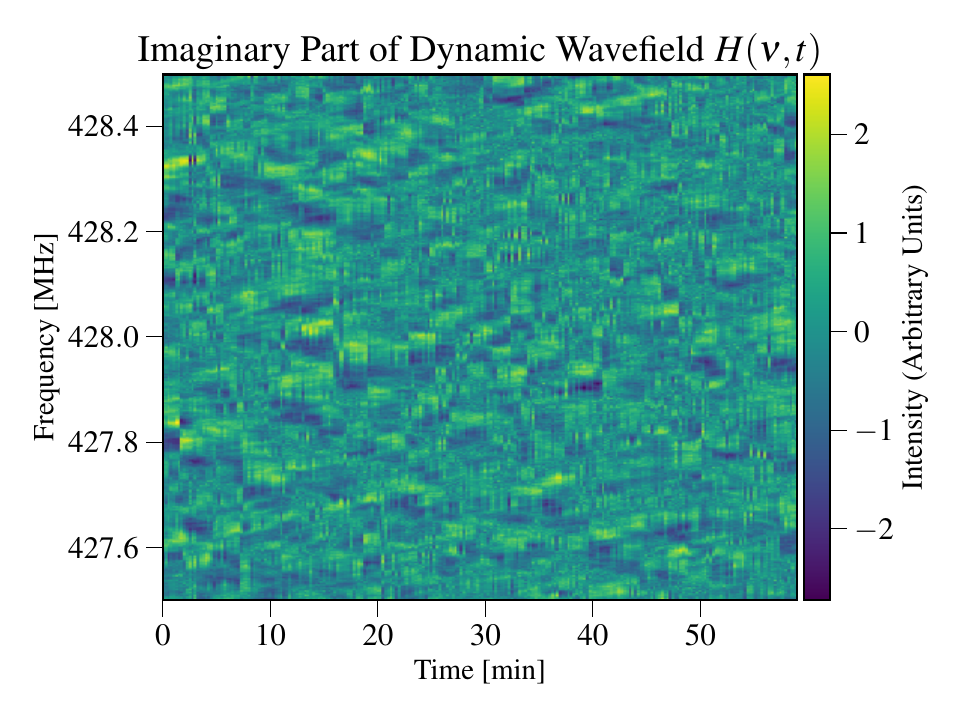} }}\quad
    \subfloat[]{ {\hspace{0cm}\includegraphics[width=0.51\textwidth]{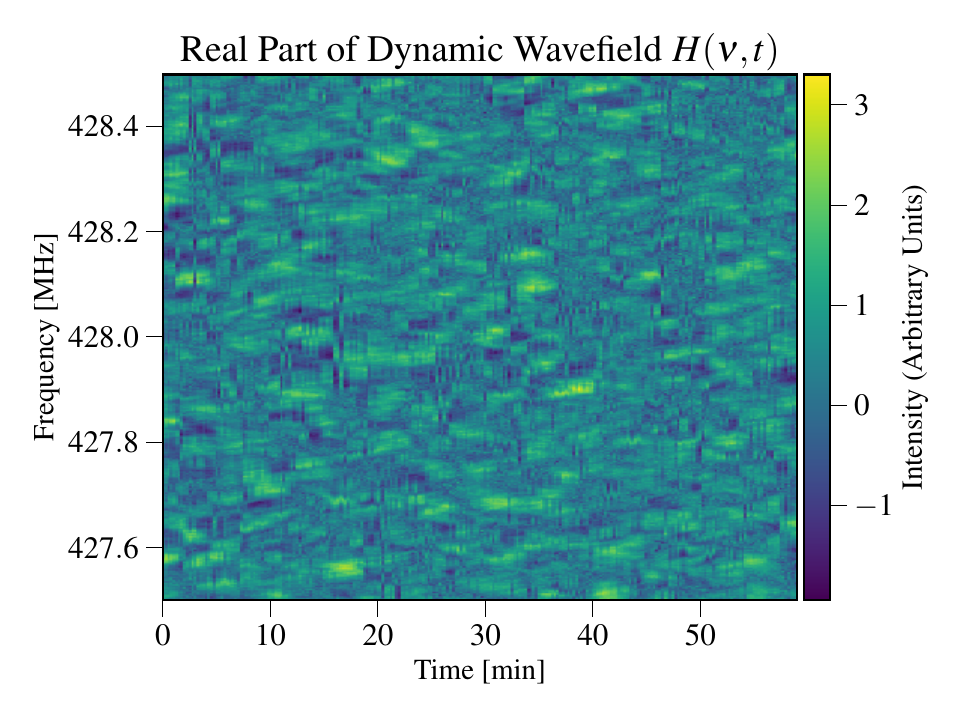} }}\\
    \subfloat[]{{\includegraphics[width=0.51\textwidth]{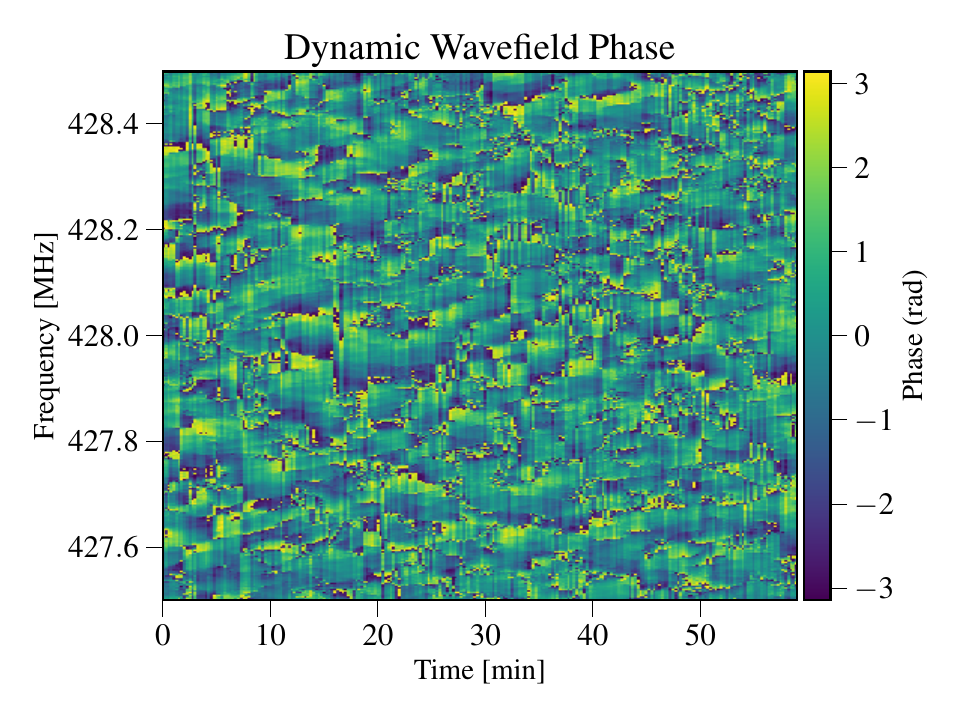} }}%
    \caption{Imaginary (top left), real (top right), and phase components of the dynamic wavefield for a PSR B1937+21 observation in the full-deconvolution regime. The prevalence of structure throughout all three images indicates this observation has complete phase retrieval.}%
    \label{full_decon_wavefield}%
\end{figure*}

\begin{figure*}[!ht]
    \centering
    \captionsetup[subfigure]{labelformat=empty}
    \subfloat[]{ {\hspace{-1cm}\includegraphics[width=0.51\textwidth]{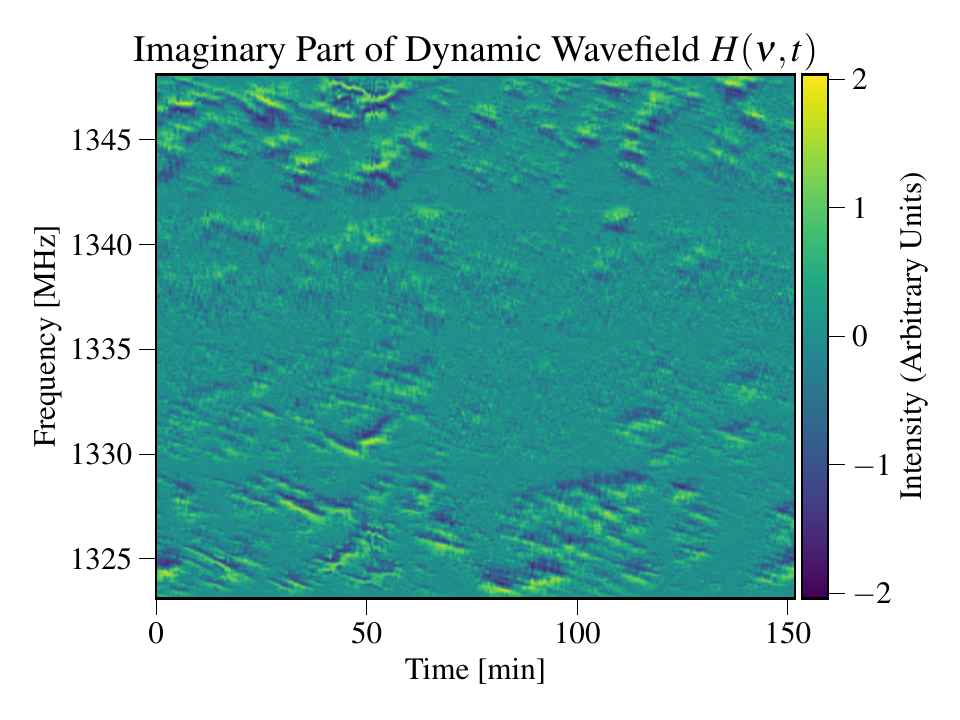} }}\quad
    \subfloat[]{ {\hspace{0cm}\includegraphics[width=0.51\textwidth]{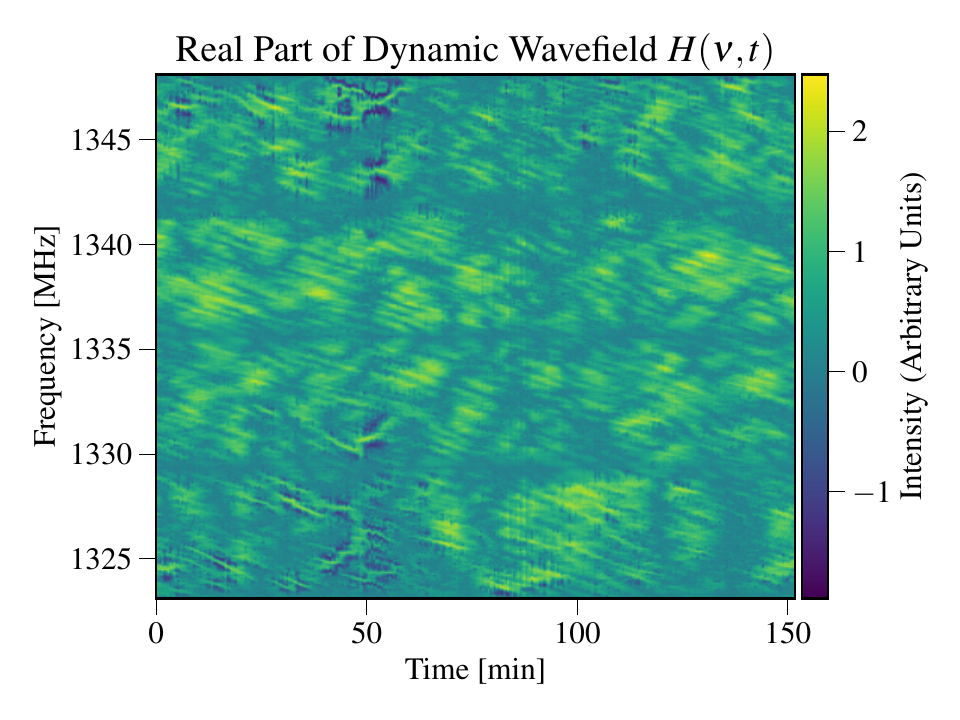} }}\\
    \subfloat[]{{\includegraphics[width=0.51\textwidth]{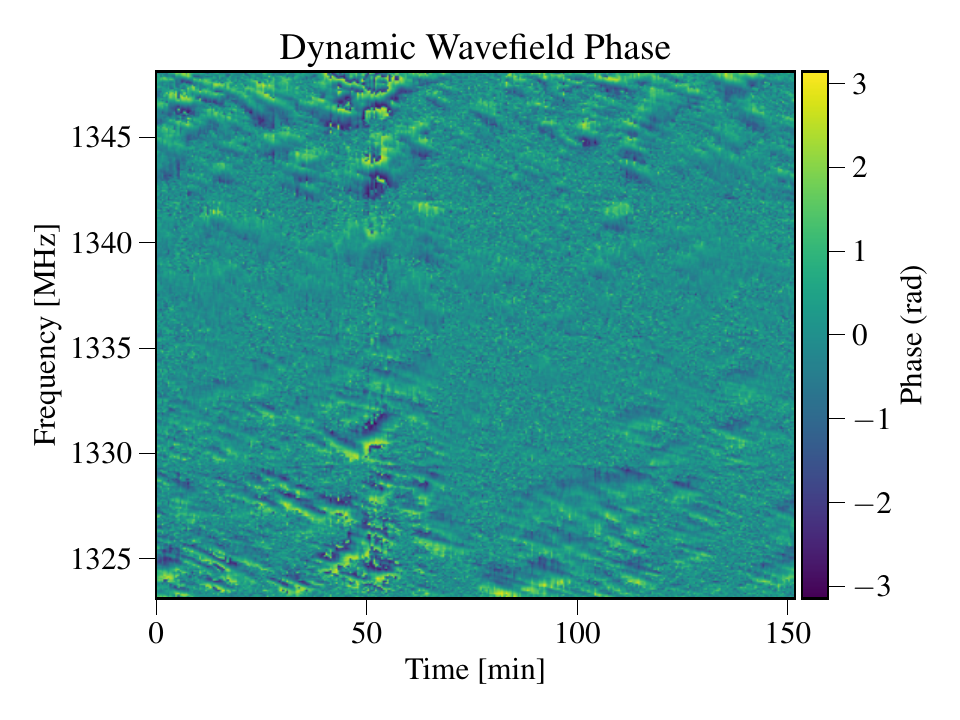} }}%
    \caption{Imaginary (top left), real (top right), and phase components of the dynamic wavefield for a PSR B1937+21  observation in the partial-deconvolution regime. The inconsistency of structure throughout all three images indicates this observation has incomplete phase retrieval, which is distinguishable as areas of uniformly distributed phase. This is in contrast with the distinct bands of fairly constant phase, which indicate complete phase retrieval. Horizontal bands in the upper right image are gaps between PFB channels.}%
    \label{non_decon_wavefield}%
\end{figure*}

\begin{figure*}[htbp]
\centering
\includegraphics[width=0.43\textwidth]{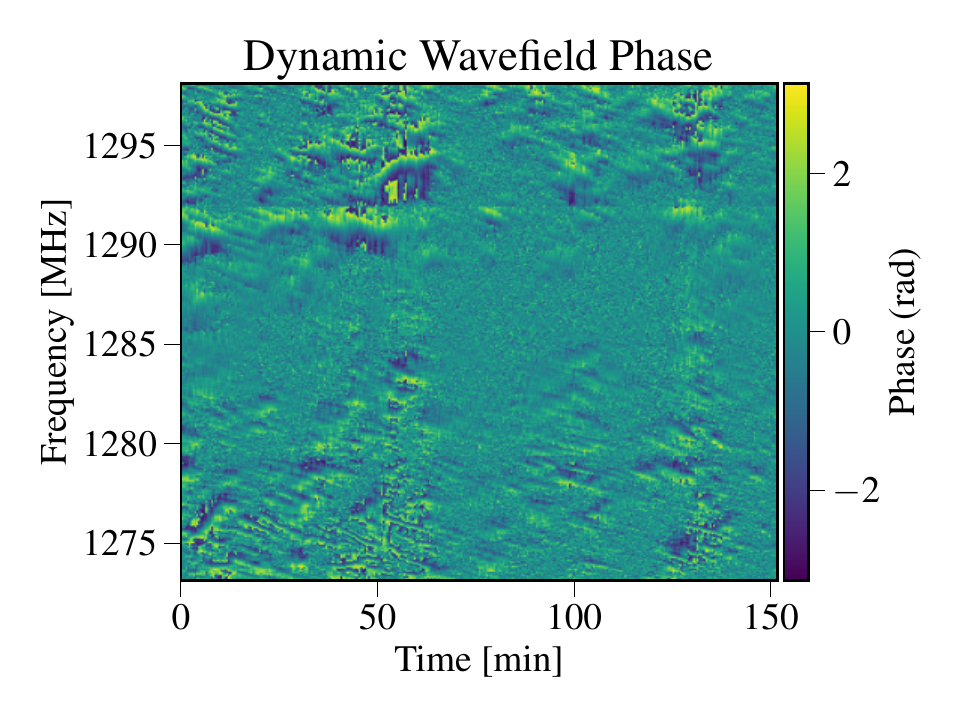}
\hfil
\includegraphics[width=0.43\textwidth]{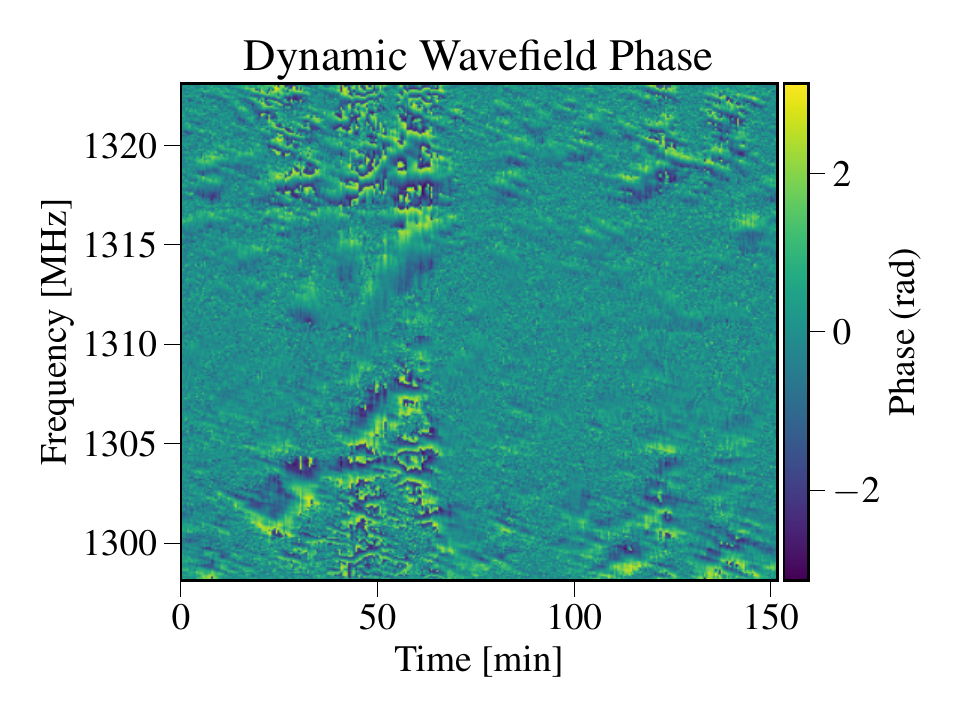}

\medskip
\includegraphics[width=0.43\textwidth]{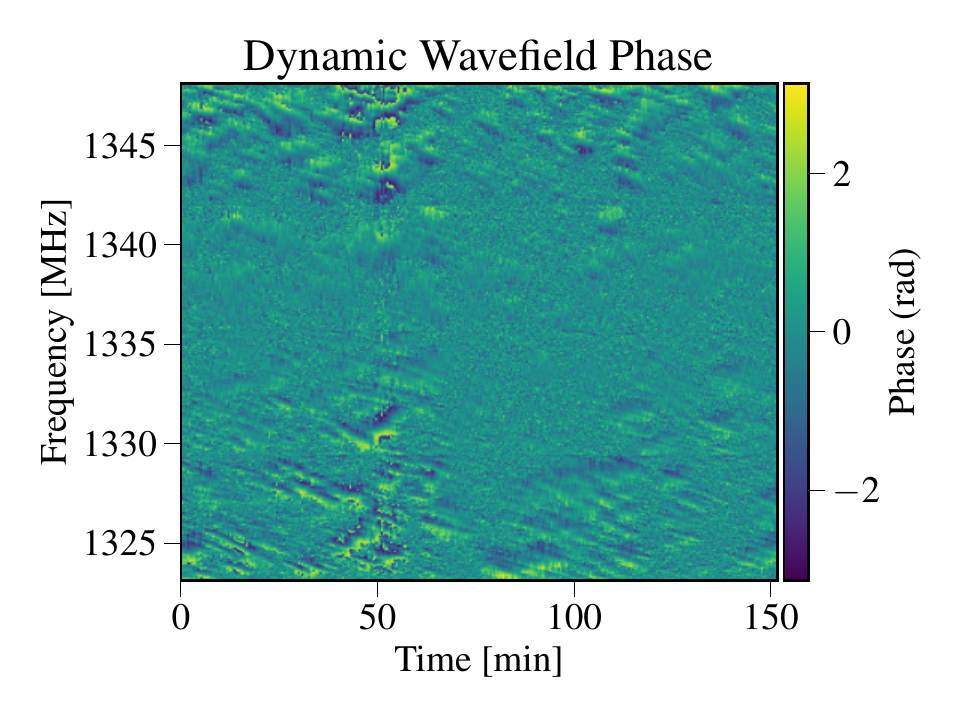}
\hfil
\includegraphics[width=0.43\textwidth]{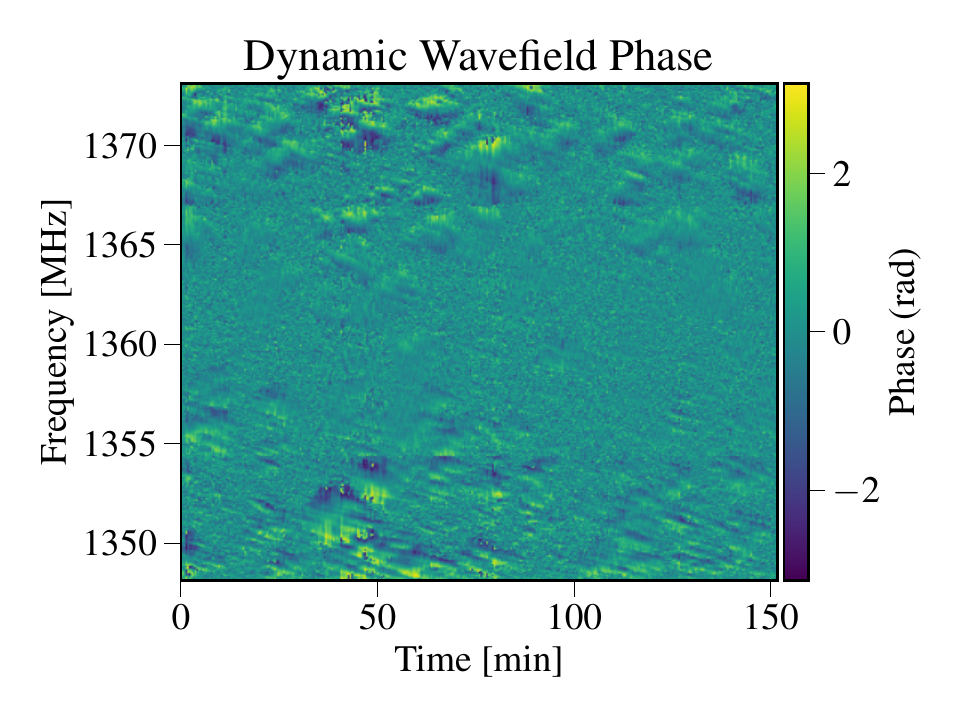}

\medskip
\includegraphics[width=0.43\textwidth]{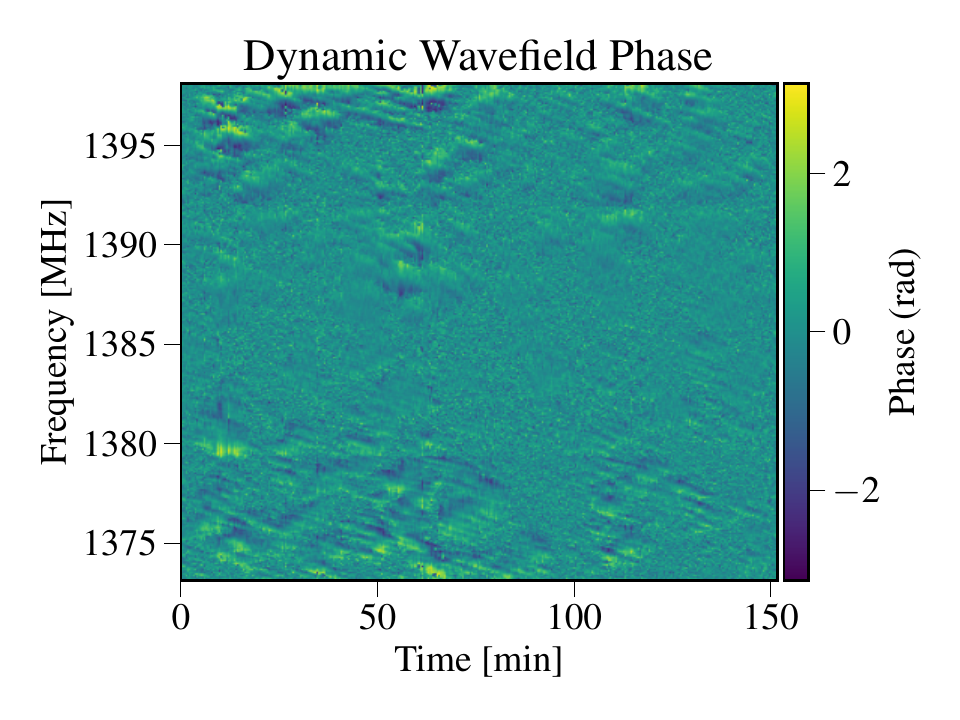}
\hfil
\includegraphics[width=0.43\textwidth]{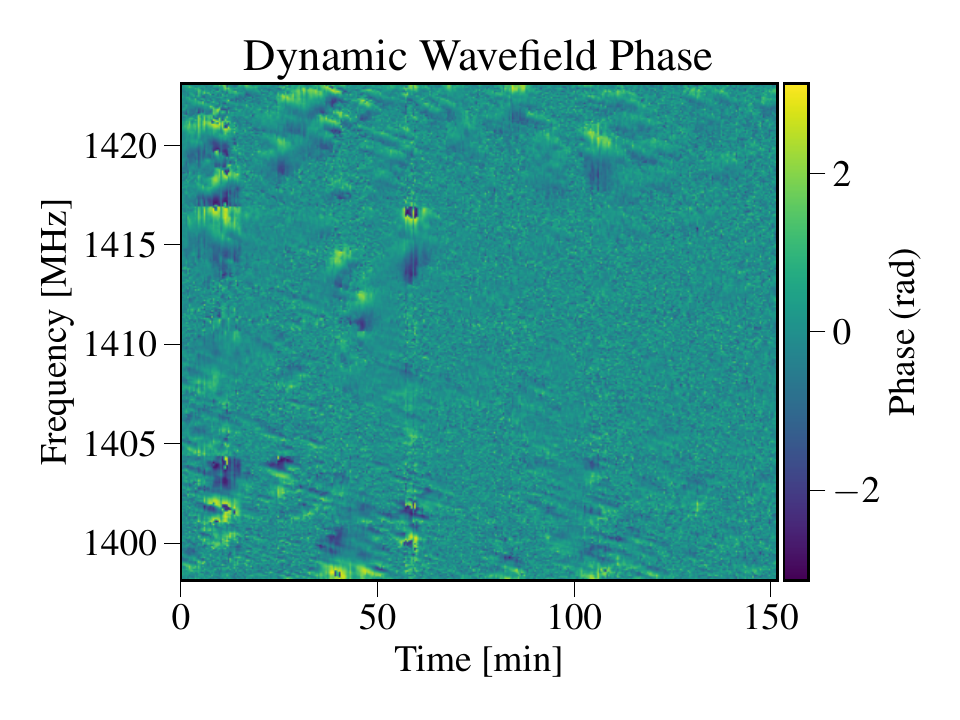}

\medskip
\includegraphics[width=0.43\textwidth]{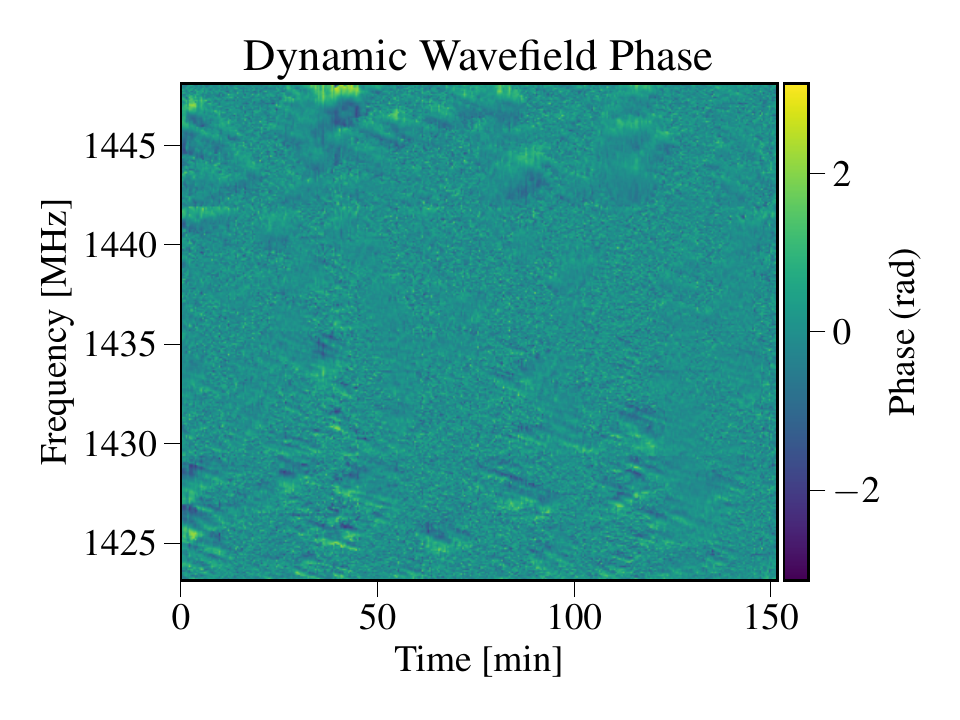}
\hfil
\includegraphics[width=0.43\textwidth]{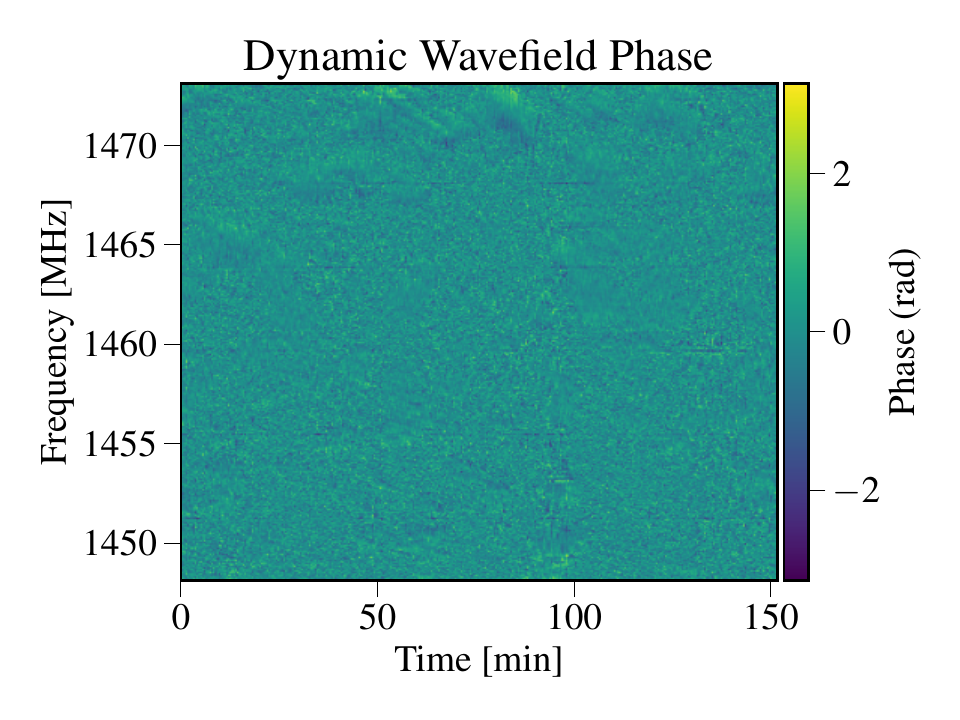}
\caption{Example dynamic wavefield phase retrieval over frequency for an observation of PSR B1937+21 in the partial deconvolution regime. As indicated by the continuity of cyclic merit, the closer to the full deconvolution regime one observes, the higher the cyclic merit and the more complete phase retrieval one obtains in their observations.}
\label{phase_freq}
\end{figure*}

Examples of imaginary, real, and phase components of dynamic wavefields can be seen in Figures \ref{full_decon_wavefield} and \ref{non_decon_wavefield} for instances of full- and partial-deconvolution, respectively. In all subsets of the full-deconvolution dynamic wavefield, there is significant and continuous structure, indicating that there is complete phase retrieval. Conversely, there are significant patches of noise in all subsets of the partial-deconvolution dynamic wavefield interspersed with patches of signal, indicating that, while phase information was recovered, phase retrieval as a whole was incomplete. Across all observations used in this paper, an examination of dynamic wavefield phases as a function observing frequency reveals a larger ratio of signal structure relative to noise the closer one gets to the full-deconvolution regime, indicating that there is a relation between observing frequency and complete phase retrieval rather than a sudden transition once the threshold between regimes has been crossed. This finding (seen in Figure \ref{phase_freq}) is in line with the continuous nature of the cyclic figure of merit described earlier, as well as what was shown in Figure \ref{merit_freq}. This indicates that significant phase retrieval can occur even outside of the full deconvolution regime, assuming a sufficient combination of high S/N and scattering delay.

\par Upon closer inspection, two significant traits emerge regarding the resolved structures in the partially-deconvolved dynamic wavefield: The first is that the recovered structures in both the dynamic wavefield phase and the imaginary component of the dynamic wavefield are concurrently very similar and located in the same regions in time-frequency space. This can largely be attributed to many of the recovered phases still being near zero, as the quantities depicted in the imaginary and phase components of the dynamic wavefield will be proportional at small angles; In a signal where either the real or imaginary part of a signal is much larger than the other, the phase will be much more sensitive to small changes in the smaller quantity. This resemblance is present across all of our data in the partial deconvolution regime, and could represent a limitation of cyclic spectroscopy in regimes where full phase retrieval is not possible. Perhaps relatedly is the real part of the dynamic wavefield bearing a strong likeness to the scintillation pattern expected in the observation's dynamic spectrum, while the imaginary component shows little, if any, resemblance. This similarity is present across all data for which we have partial phase retrieval, and may, again, simply be due to the imaginary component of our partially recovered signal being much smaller than the real component. In general, the amplitude pattern of the dynamic wavefield will always resemble the structure seen in the corresponding dynamic spectrum, which itself only contains amplitude information.

\par
\begin{figure*}[!ht]
    \centering
    \captionsetup[subfigure]{labelformat=empty}
    \subfloat[]{\hspace*{-.8cm} {\includegraphics[width=0.55\textwidth]{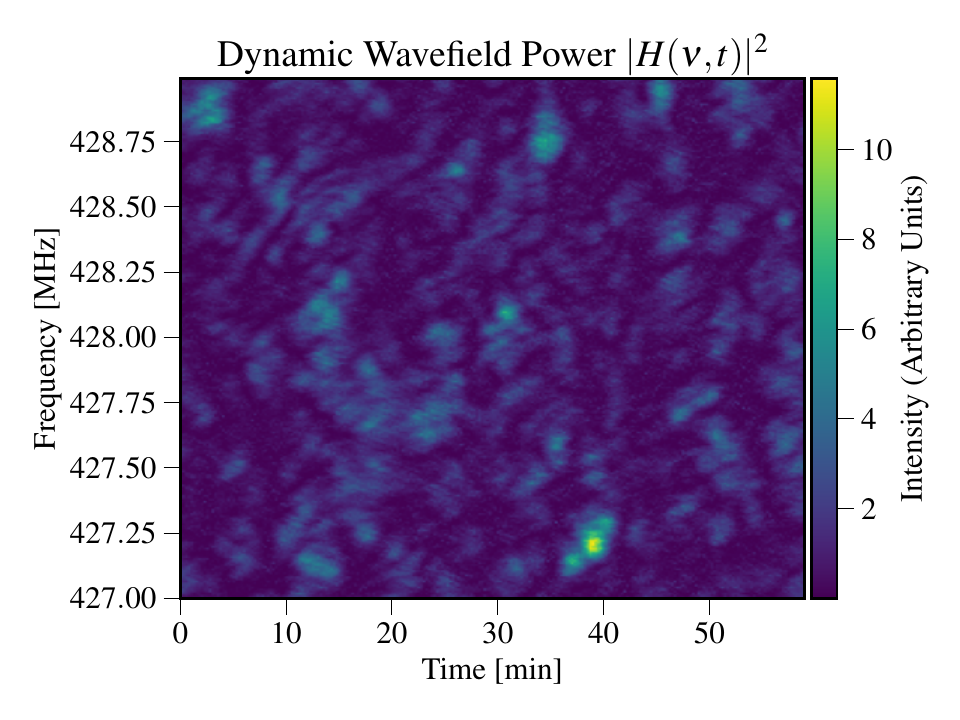} }}%
    \subfloat[]{{\includegraphics[width=0.55\textwidth]{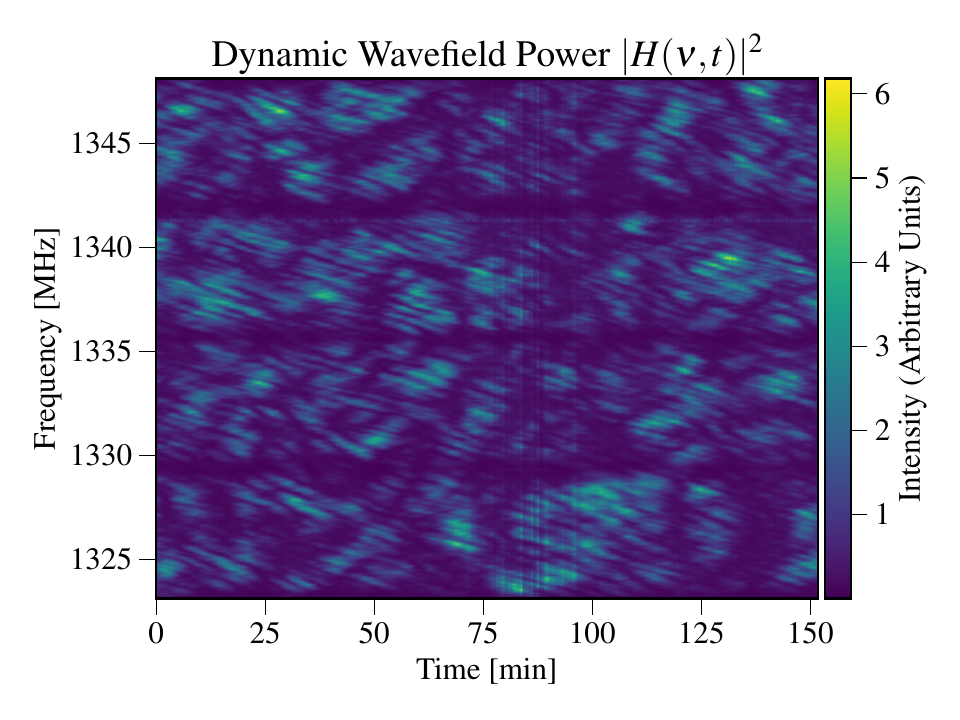} }}%
    \caption{Dynamic wavefield powers from PSR B1937+21 observations with full (left) and partial (right) phase retrieval. As indicated by the full reconstruction of the scintillation pattern, the dynamic wavefield power with incomplete phase retrieval still recovers the frequency response intensity of the ISM due to this quantity relying on signal amplitude. Horizontal bands in the image on the right are gaps between PFB channels.}%
    \label{wavefield_power}%
\end{figure*}

 This similarity can be seen by examining the dynamic wavefield power, which by definition is also an amplitude-only function. As shown in Figure \ref{wavefield_power}, despite the incomplete phase retrieval, the dynamic wavefield power appears to have complete signal recovery and strongly mirrors the frequency response pattern one would expect in the dynamic spectrum. However, while dynamic spectra are created exclusively from amplitude information, the additional phase information in the signals that comprise the dynamic wavefield power makes the latter data product more representative of the interaction between the ISM and the intrinsic pulse emission. The dynamic spectrum itself can be thought of as the dynamic wavefield power at the $0^{\rm th}$ cyclic harmonic \citep{wdv13}.

\par
\begin{figure}[!ht]
    \centering
    \captionsetup[subfigure]{labelformat=empty}
    {\hspace*{-.8cm}\includegraphics[width=0.55\textwidth]{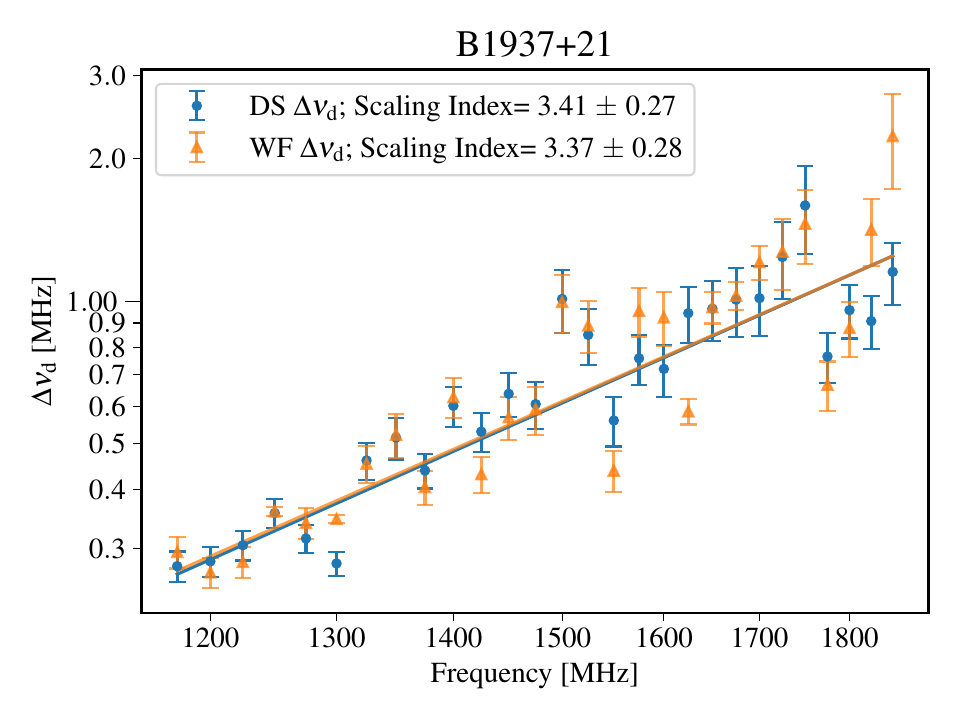} }%
    \caption{Scintillation bandwidth power law fits using both dynamic spectra (DS; blue circles) and dynamic wavefield power (WF; orange triangles) from an observation of PSR B1937+21. Any differences in measurements between the two trends can primarily be attributed to the difference in S/N of the two approaches, and possibly due to the additional harmonic information in the dynamic wavefield power.}%
    \label{power_law_wf}%
\end{figure}

 For these reasons, analyses that are typically accomplished using dynamic spectra can also be accomplished with the dynamic wavefield power, even in the partial deconvolution regime, but with the added benefit of partial phase retrieval as well as information contained in the higher harmonics of the data. A valuable example of this are efforts to constrain electron density wavenumber spectra and the resulting turbulence along various lines of sight through the galaxy, which can be accomplished by performing frequency-dependent power law fits of scintillation bandwidth or timescale.  Such studies are ideal for pulsars that stand to benefit the most from cyclic spectroscopy, as highly scattered pulsars have more scintles in a given span of observing frequency and time, and, consequentially, smaller uncertainties on scintillation measurements. Such scenarios allow for tightly constrained fits, especially when performed over comparatively wide observing bandwidths. When we perform power law fits of the form $\Delta \nu_{\rm d} = a\nu^{x}$ using both dynamic spectra and dynamic wavefield power, we encounter almost identical trends from nearly identical measurements, as shown in Figure \ref{power_law_wf}. This result should not be too surprising, as both the dynamic spectrum and dynamic wavefield power are a consequence of the time-evolving frequency response of the ISM, but it serves as a strong demonstration of what can be accomplished with cyclic spectroscopy derived data products in the partial deconvolution regime. 

\subsection{Secondary Spectra \& Wavefields}
\begin{figure*}[!ht]
    \centering
    \captionsetup[subfigure]{labelformat=empty}
    \subfloat[]{ {\hspace{-1cm}\includegraphics[width=0.51\textwidth]{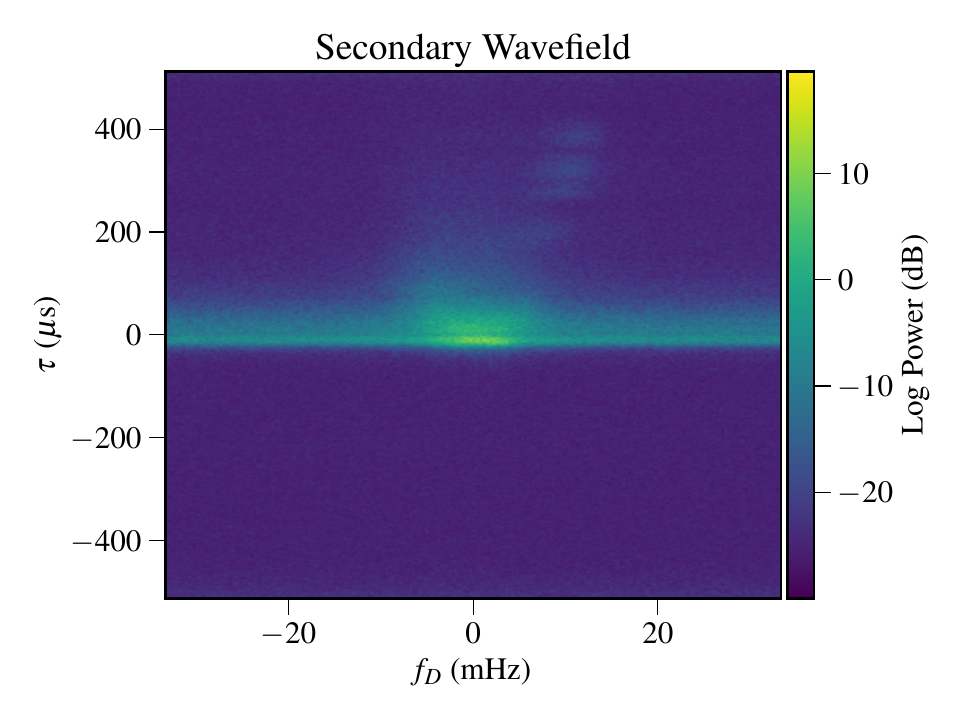} }}\quad
    \subfloat[]{ {\hspace{0cm}\includegraphics[width=0.51\textwidth]{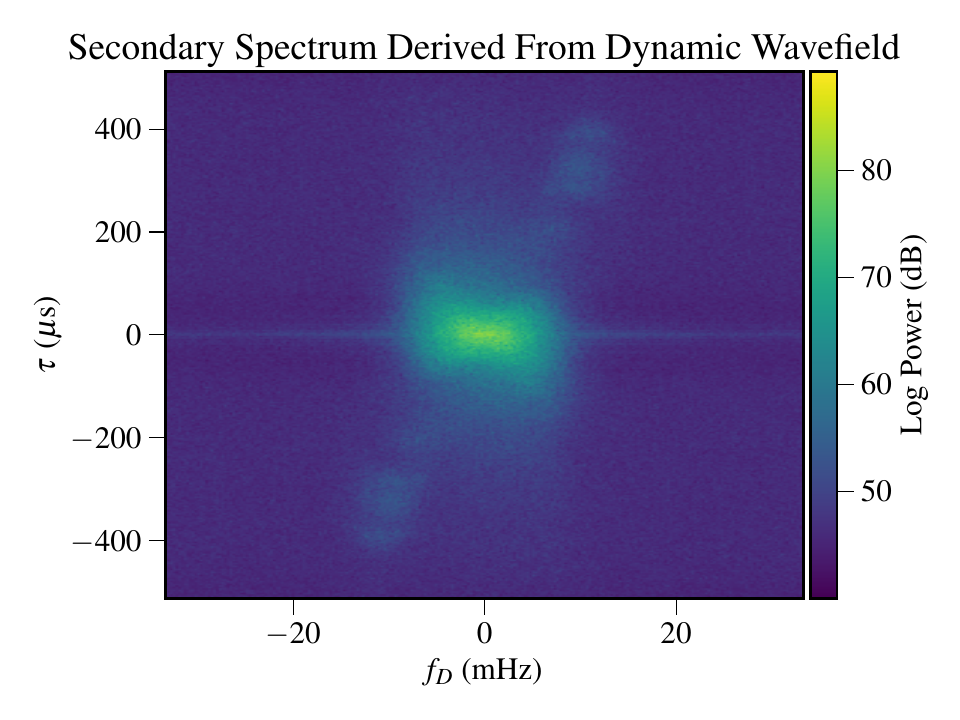} }}\\
    \subfloat[]{ {\hspace{-1cm}\includegraphics[width=0.51\textwidth]{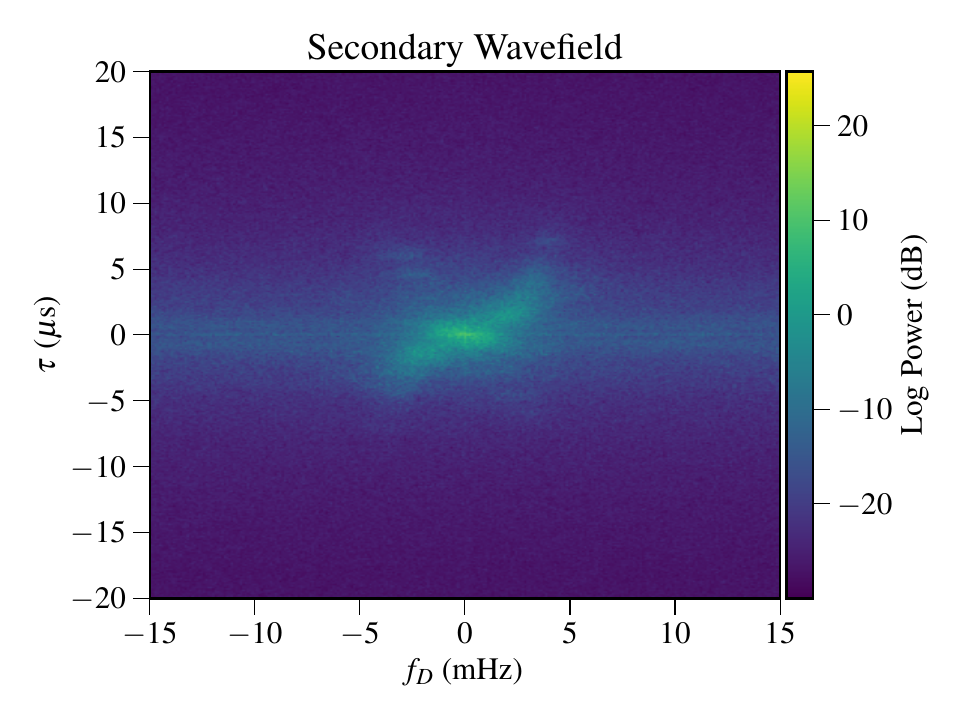}}}\quad
    \subfloat[]{{\includegraphics[width=0.51\textwidth]{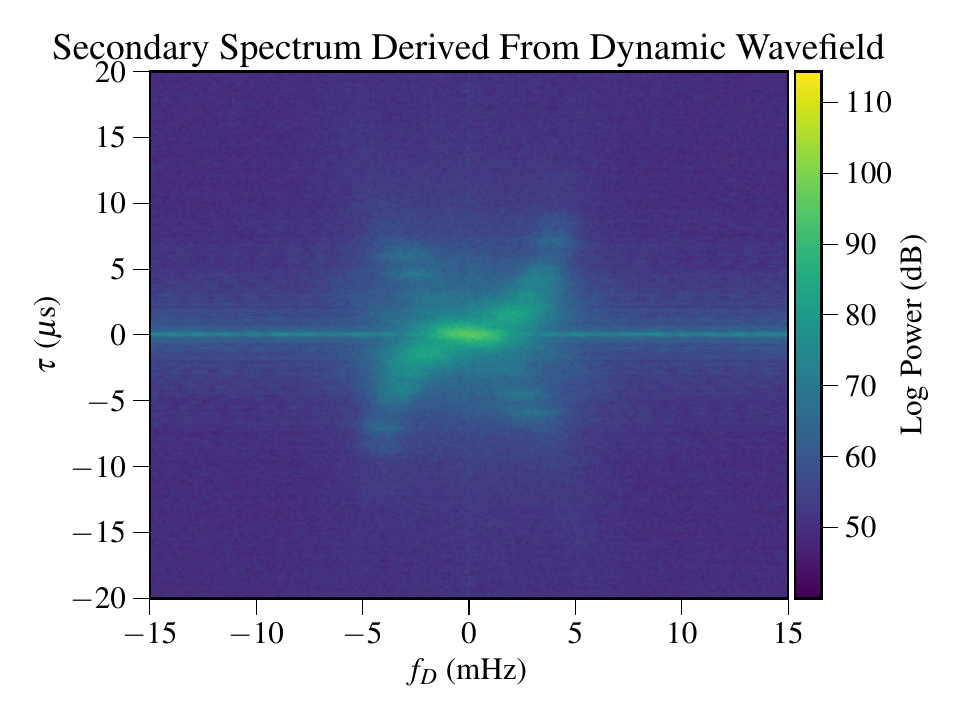}}}%
    \caption{Secondary wavefields (left) and secondary spectra (right) from PSR B1937+21 observations in the full (top) and partial (bottom) deconvolution regimes. The limited degree to which scattered images are recovered, along with the presence of images at non-physical delays in the bottom secondary wavefield indicate incomplete phase retrieval.}%
    \label{secondary_compare}%
\end{figure*}

Example secondary wavefields and spectra from observations taken in full- and partial-deconvolution can be seen in Figure \ref{secondary_compare}. While discrete scattered images in the full-deconvolution secondary spectrum translate well into flattened and easily distinguishable images of comparable S/N in the corresponding secondary wavefield, similar discrete images in the partial-deconvolution secondary spectrum are not resolved as finely or distinctly in the corresponding secondary wavefield. An additional scintillation arc seen in the secondary spectrum may be partially resolved on the right side of the secondary wavefield, although it is quite faint. Furthermore, while there is no discernable signal at negative delays in the full-deconvolution secondary wavefield (which makes sense, given that such delays are unphysical; \citealt{wdv13}), there is a fair bit of signal at negative delays in the partial-deconvolution secondary wavefield, caused partially by incomplete phase retrieval. However, unlike the corresponding secondary spectrum, these structures are not symmetrical with features at positive delays. This is likely due in some measure to the partial recovery of the phase, but at least one factor that even mathematically allows for the presence of this structure is that, unlike the secondary spectrum, which is created from the Fourier transform of a real-valued function, the secondary wavefield is derived from the Fourier transform of a complex function, and so is not guaranteed the same symmetry. 
\par Although complete recovery of the secondary wavefield is not possible in the partial-deconvolution regime, techniques such as the $\theta-\theta$ transform (see discussion earlier in Section \ref{intro} for a brief summary of the technique) can serve a complementary role in phase retrieval, as it can be used to reconstruct the dynamic and secondary wavefield \citep{baker_2021}. However, one significant caveat to this method is that it only operates as a suitable substitute in instances of individual phase screens with highly anisotropic scattering \citep{theta_theta}. Many of the most highly scattered pulsars, which constitute the majority of sources for which these phase retrieval techniques are valuable, often exhibit much more complicated lines of sight. For example, our observations of PSR B1937+21 display significant deviations from highly anisotropic scattering and exhibit multiple phase screens. Since $\theta-\theta$ first requires an estimate of arc curvature from the secondary spectrum, scintillation arcs at significantly different curvatures may be distorted in the resulting $\theta-\theta$ spectrum and, consequently the secondary wavefield. 
\par
\begin{figure*}[!ht]
    \centering
    \includegraphics[scale = 0.6]{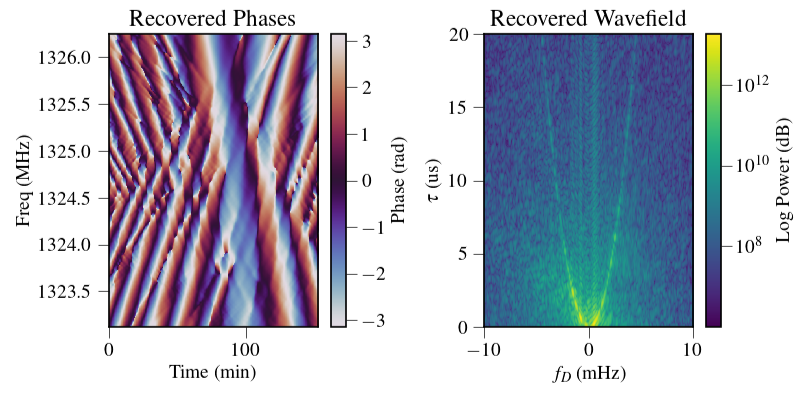}
    \caption{Attempted recovered phase (left) and secondary wavefield (right) using the $\theta-\theta$ implementation found in \textsc{scintools} \citep{scintools, theta_theta, baker_2021} of the partial-deconvolution regime observation of PSR B1937+21 shown in earlier figures. Assumptions of anisotropic, single-screen scattering inherent to the $\theta-\theta$ transform may bias secondary wavefield recovery for pulsars with more complicated lines of sight.}
    \label{theta_attempt}
\end{figure*}

 This bias can be seen in see Figure \ref{theta_attempt}, where we show our attempted $\theta-\theta$ secondary wavefield reconstruction of the partial-deconvolution regime observation shown in Figure \ref{secondary_compare}. While the asymmetry in the brightness distribution of the primary arc appears to match what is seen in the secondary spectrum, the isolated scattered images are insufficiently recovered; unlike the cyclic spectrum-derived secondary wavefield for this observation, the locations of the islands of power do not generally match those in the secondary spectrum, or islands of power in the secondary spectrum are not present in the $\theta-\theta$ secondary wavefield. Additionally, due to the inherent uncertainty in phase retrieval on data taken with single dish instruments, such as the above-mentioned Arecibo Observatory observation, $\theta-\theta$ may not be able to fully recover phases without the aid of setups like very long baseline interferometry \citep{baker_2023}.
\break
\subsection{Dynamic Spectra}
Through our use of cyclic spectroscopy, we gain benefits even in data products that can be already be acquired through standard Fourier spectroscopy. Since all of our L--band data have both 6.1 kHz frequency resolution and 1024 pulse phase bins, they can be used for pulsar timing, studies of the interstellar medium, and tracking of pulse time-of-arrival delays caused by interstellar scattering. With NANOGrav's current frequency resolution of approximately 1.5 MHz, we can only resolve scattering delays up to around 33 ns, assuming at least three frequency channels per scintle is deemed sufficient to claim a measurement \citep{turner_scat}. 
\par
\begin{figure*}[!ht]
    \centering
    \captionsetup[subfigure]{labelformat=empty}
    \subfloat[]{\hspace*{-.8cm}{\includegraphics[width=0.55\textwidth]{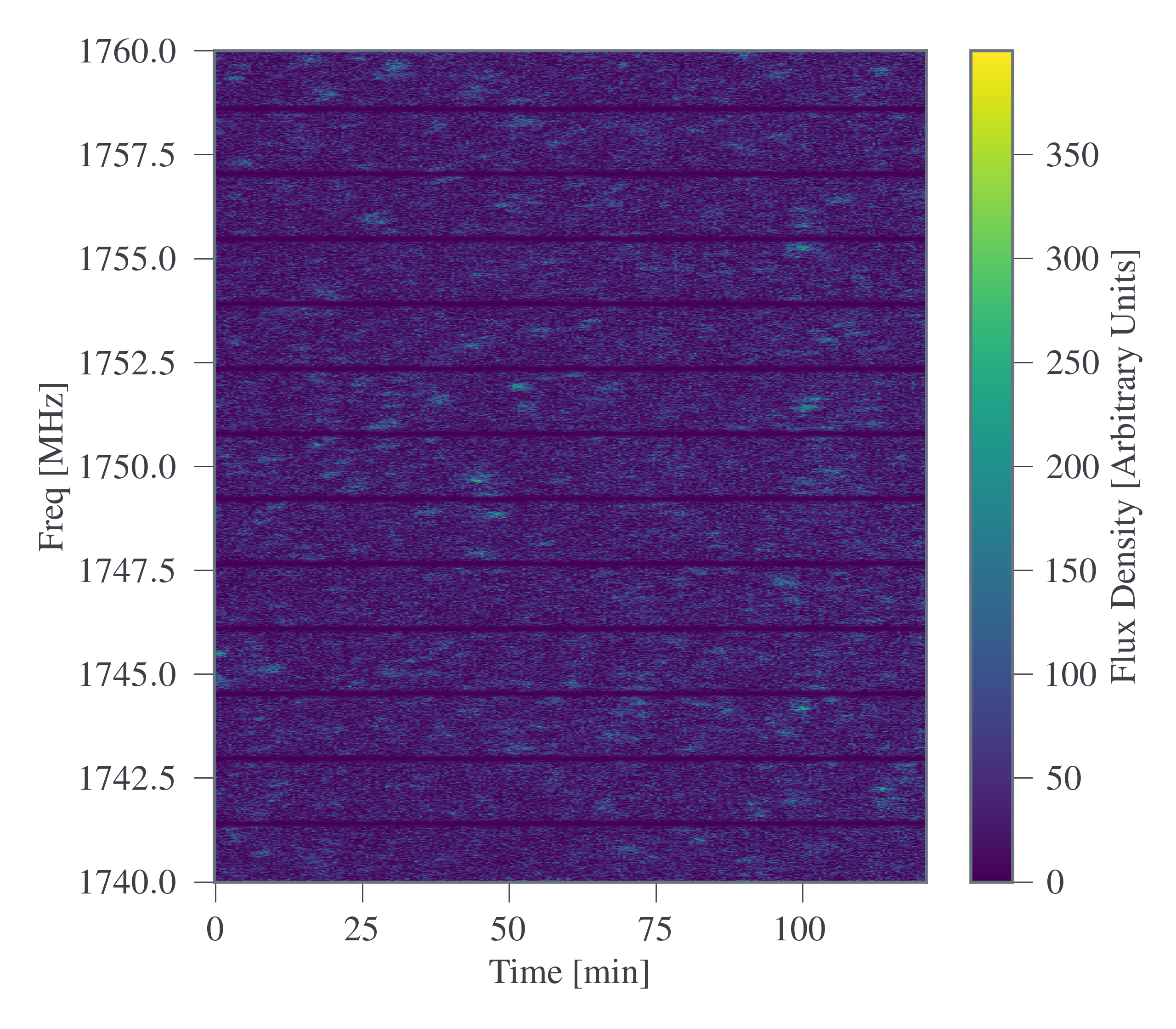}}}
    \subfloat[]{{\includegraphics[width=0.55\textwidth]{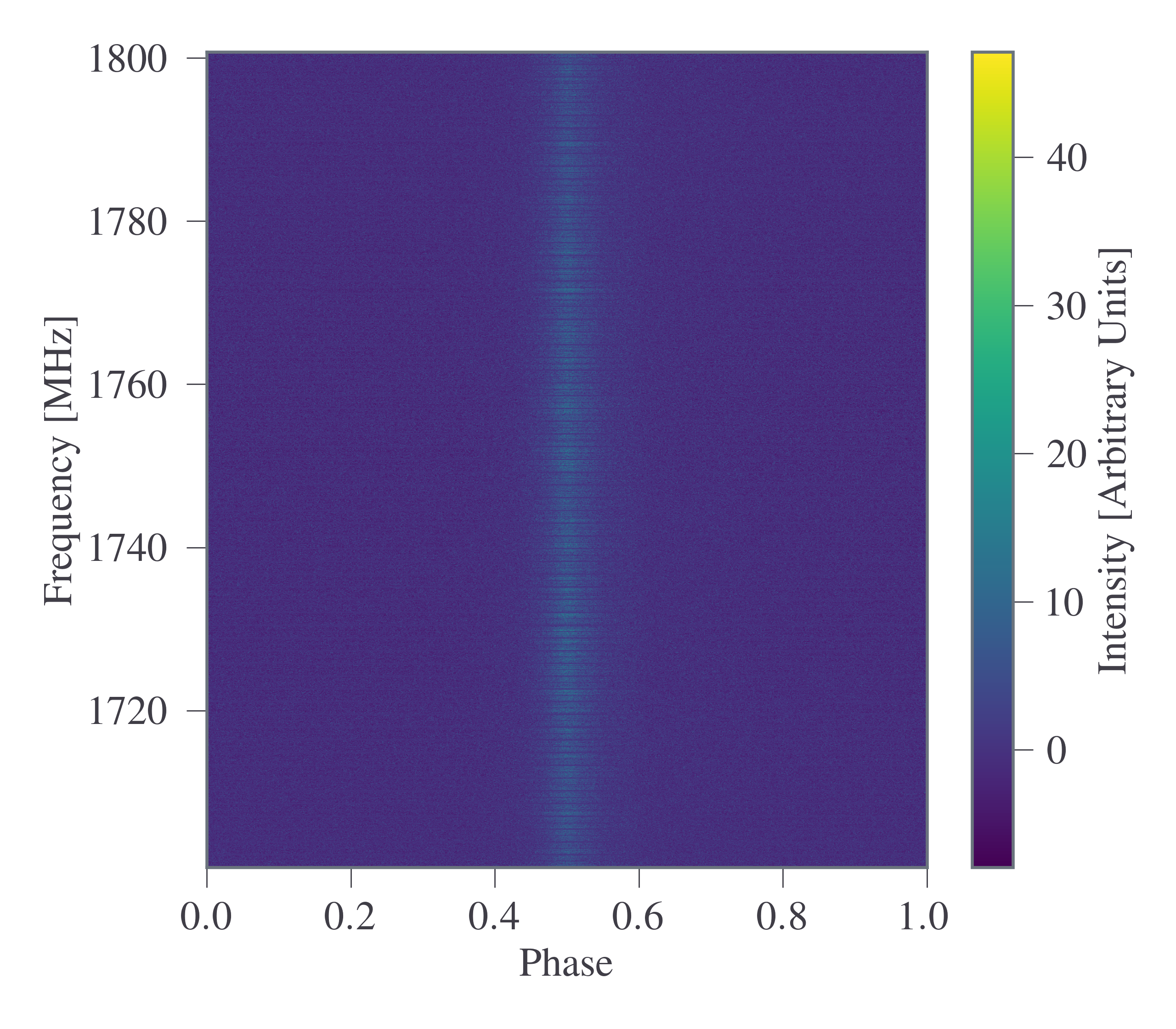} }}%
    \caption{Recovered dynamic spectrum (left) and intensity pulse profile (right) of PSR J1643--1224 from an observation processed with cyclic spectroscopy. The periodic black bands indicate gaps between polyphase filterbank channels.}
    \label{1643_ex}
\end{figure*}

 In the latter scenario, pulsars like PSR J1643--1224, whose timing residuals are likely dominated by scattering, and have been observed to exhibit scattering delays on the order of a few microseconds at L--band \citep{main_leap}, would never have resolvable scintles in NANOGrav data. Consequently, the only quantifiable estimate NANOGrav could make using the scintillation bandwidth approach is placing lower limits of 33 ns, potentially two orders of magnitude lower than its true delay in a given observation. Otherwise, they would have to resort to techniques such as modeling via timing or the pulse profile. Conversely, in our data, both this pulsar's scintles and pulse profile are fully resolved, as can be seen in Figure \ref{1643_ex}. Additionally, with this setup, even if the S/N in a dynamic spectrum is too low at the full pulse phase and frequency resolution used, one can simply bin average the profile. 
\par Of particular note in this dynamic spectrum are the periodic black bands, which correspond to gaps between the polyphase filterbank channels. These gaps need not be inherent to all observations, although mitigating their presence would require developing instruments that provide overlapped filterbank channels or developing processing software that allows for the combination of data from adjacent channels. For data in which these bands are present, even though cyclic spectroscopy allows one to upsample their data with an arbitrary number of cyclic channels per filterbank channel, additional care needs to be taken when creating observing configurations. Allowing an observation to have too many filterbank channels could result in filterbank gaps that intersect individual scintles, even if the scintles themselves require cyclic channelization to be resolved. Conversely, while minimizing the number of filterbank channels would eliminate gaps, the additional number of cyclic channels per filterbank channel required to achieve the desired frequency resolution would significantly increase the time required to process the observation, even on a system like Green Bank Observatory's forthcoming cyclic spectroscopy backend. Relatedly, while one could, in theory, request an observation in which only one polyphase filterbank channel spans their entire observing bandwidth, not only would such a setup be extremely computationally expensive to process with a sufficient number of cyclic channels, most telescopes do not even allow for such configurations. For these reasons, it is crucial for observers to determine a satisfactory medium between their desired post-cyclic processing frequency resolution and the configurations available within existing pulsar observing backends. 
\par
\begin{figure*}[!ht]
    \centering
    \captionsetup[subfigure]{labelformat=empty}
    \subfloat[]{\hspace*{-.8cm}{\includegraphics[width=0.55\textwidth]{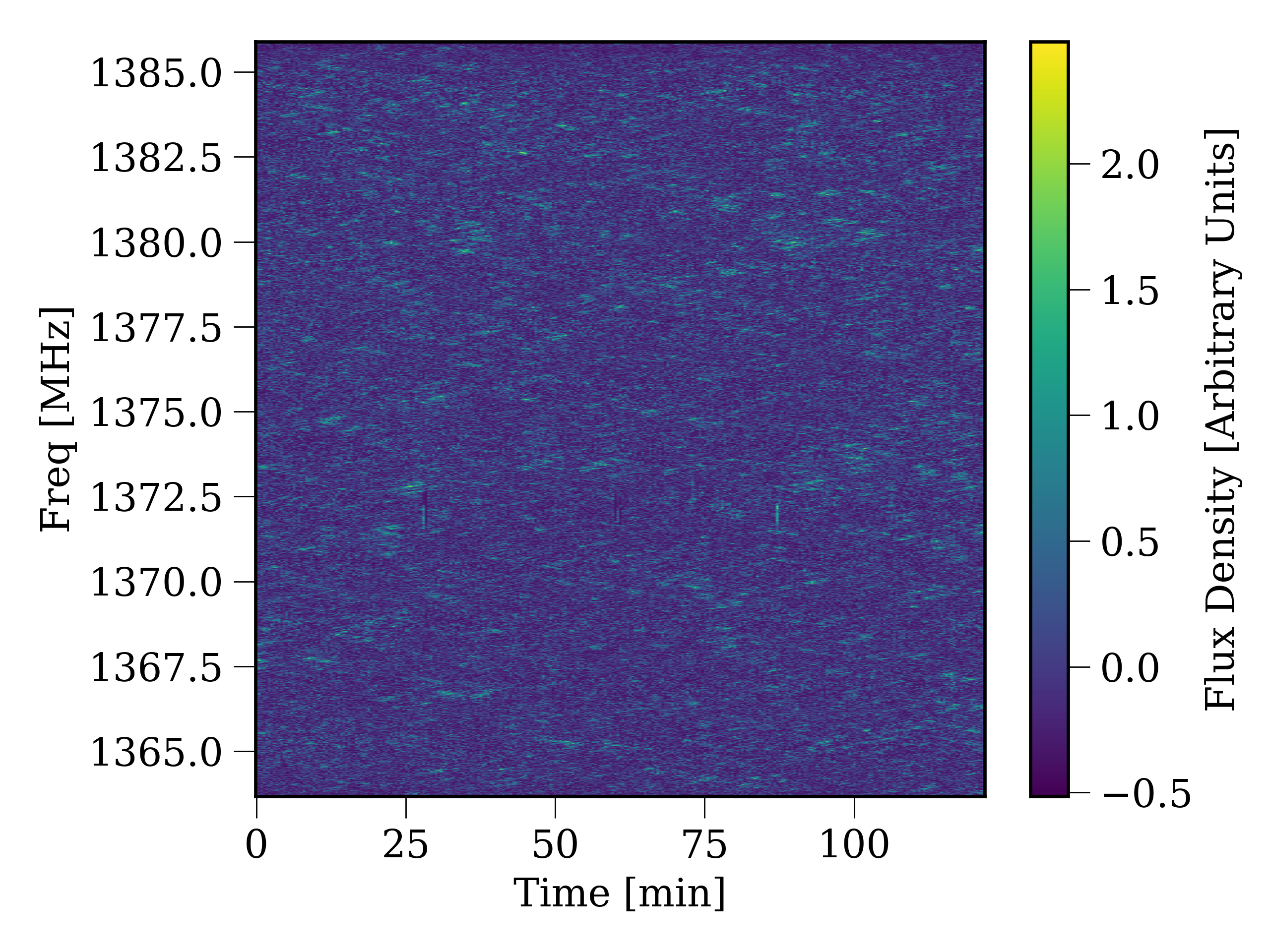} }}%
    \subfloat[]{{\includegraphics[width=0.55\textwidth]{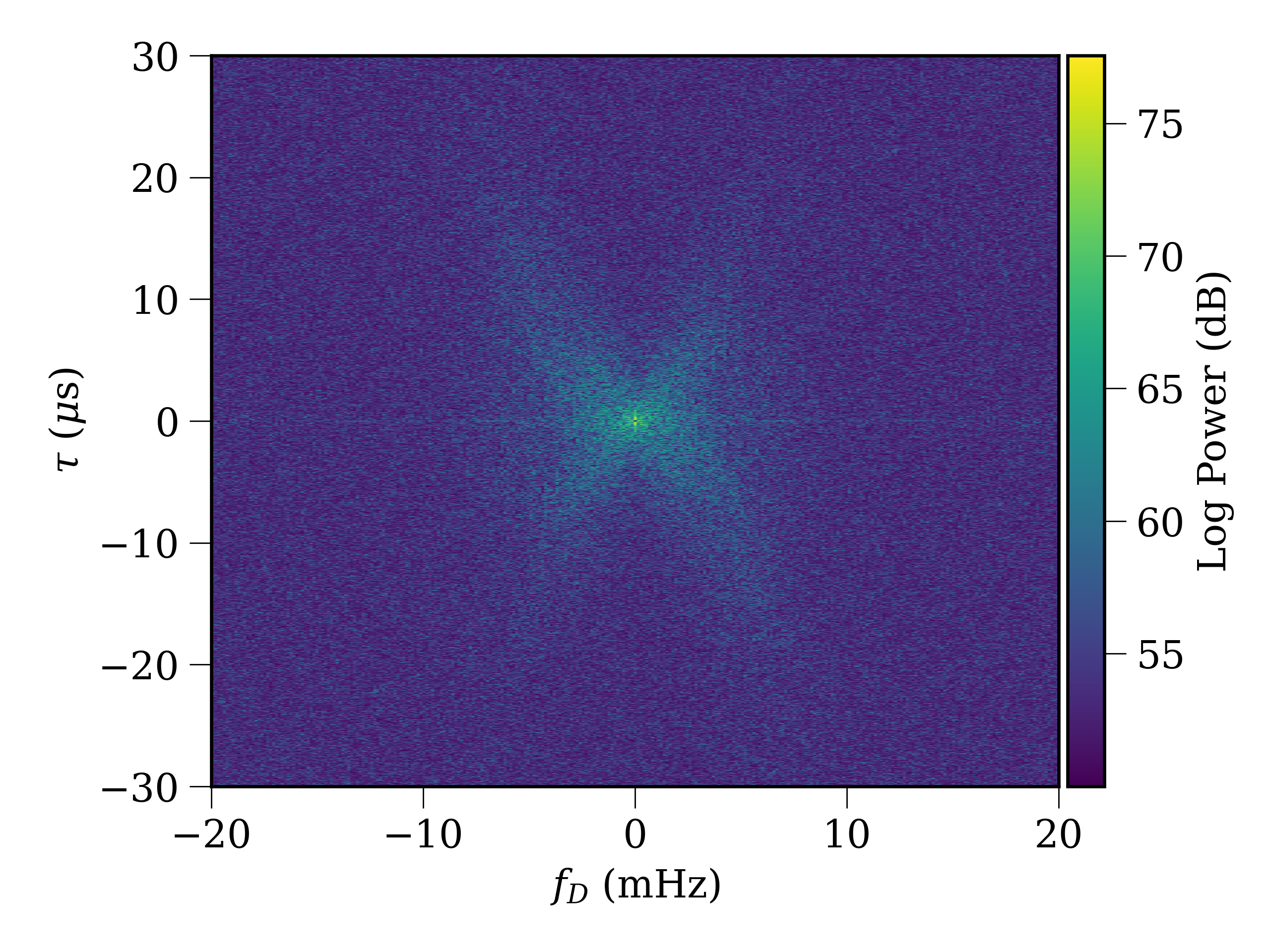} }}%
    \caption{Example dynamic (left) and secondary (right) spectra for PSR J1643--1224 across a single PFB channel. Significant, detailed scintillation structure is visible in the dynamic spectrum and a scintillation arc is visibile in the secondary spectrum.}%
    \label{dyn_and_sec}%
\end{figure*}

 That being said, while these filterbank gaps present a nuisance, one can still work around these constraints and perform high-quality science in the proper configuration, particularly for highly scattered pulsars where lots of scintles will be visible in a small range of observing frequencies. These small frequency ranges are in fact desirable to avoid significant evolution in scintle size across the observing band, especially at lower frequencies. Zooming in on a single polyphase filterbank channel in an observation of PSR J1643--1224 (Figure \ref{dyn_and_sec}), we can see highly detailed scintillation structure across approximately 25 MHz of observing bandwidth, and can resolve a scintillation arc in the corresponding secondary spectrum. 
\par
\begin{figure}[!ht]
    \centering
    \hspace*{-1cm}                                                           
    \includegraphics[scale = 0.6]{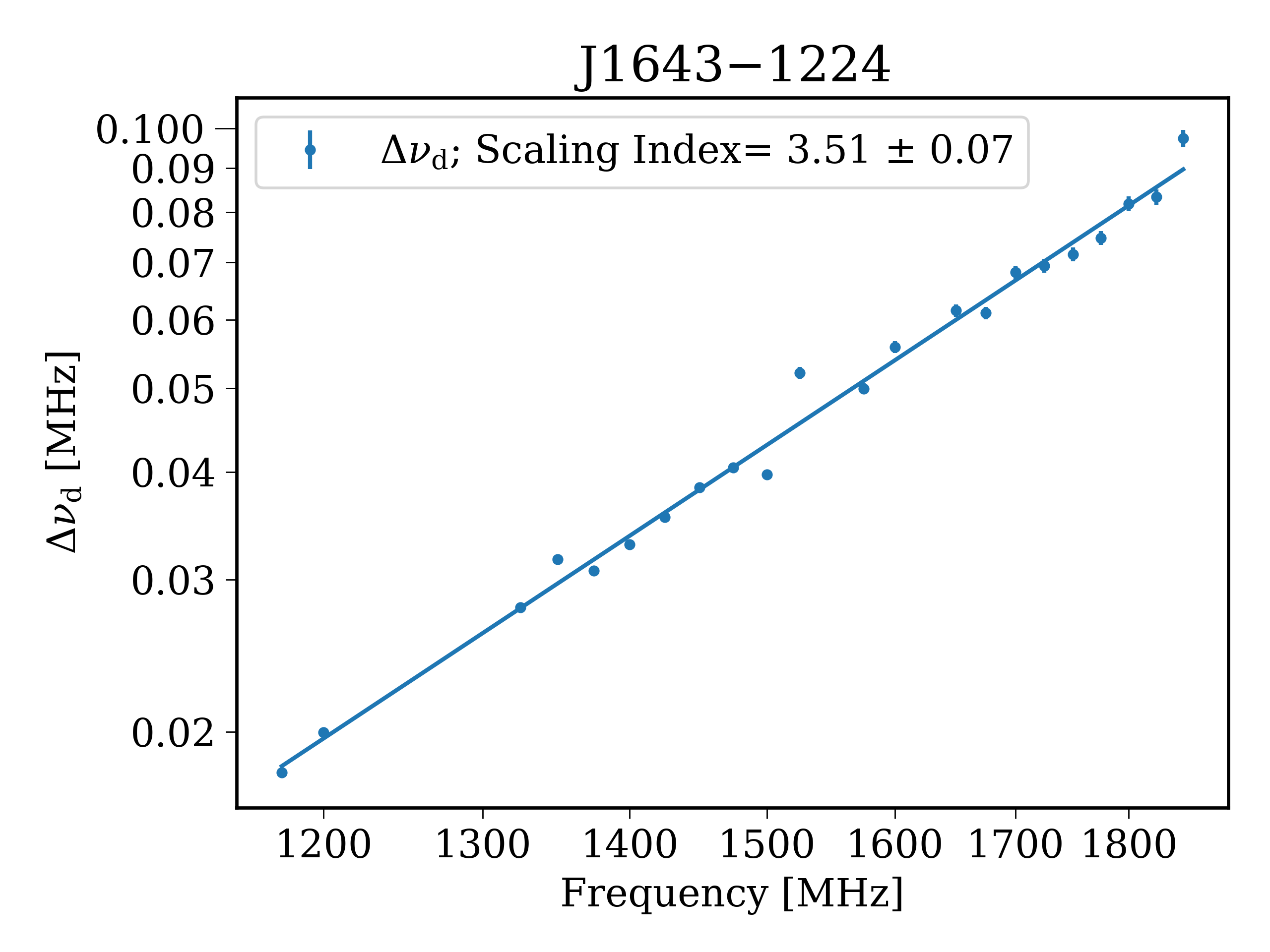}
    \caption{Scintillation bandwidth power law fit for an observation of PSR J1643$-$1224. The narrow scintles allow for precise estimations and many measurements across the observing band, resulting in highly constrained scaling indices. }
    \label{1643_power_law}
\end{figure}

 Similar to our discussion surrounding Figure \ref{power_law_wf}, this pulsar's incredibly narrow scintillation bandwidth, as is the case with almost all pulsars that benefit from cyclic spectroscopy, also makes it an ideal candidate to perform highly constrained estimations of turbulence along its LOS. As shown in Figure \ref{1643_power_law}, these narrow scintles result in small errors for individual measurements, and the large number of measurements that can be made across a single dynamic spectrum allow for stringent fits to scaling indices.  
\section{Discussion}
\label{sec:discussion}
\par After examination of various data products created with cyclic spectroscopy in the partial-deconvolution regime, it appears reasonable to prefer cyclic spectroscopy-processed data to Fourier spectroscopy-processed data in all regimes, assuming baseband data or a cyclic spectroscopy backend are available. Regardless of whether deconvolution is complete, there will always be additional phase information in one's data that is not present in Fourier spectroscopy-processed data. It is therefore valuable to treat data with partial deconvolution as incomplete rather than incorrect, and approach utilization from a perspective of how to take advantage of the information that is there. 
\par A prime example is the partially recovered dynamic wavefield, which allows for detailed studies of the interstellar medium due to the dynamic wavefield power both being generally recoverable and exhibiting the same scintillation structure as the dynamic spectrum, while also allowing for partial exploration of the secondary wavefield, in addition to full exploration of the secondary spectrum. More frequency channels would also mean less data needing to be excised due to narrowband RFI contamination, although even excision may not always be necessary, as RFI that does not oscillate at multiples of the inverse pulse period should not appear in the dynamic wavefield \citep{wdv13}. Even in cases where RFI is still present, one can try different trial periods and excise RFI as it is found in successive cyclic spectra. Additionally, the finer frequency channelization means that scintillation studies of the interstellar medium towards highly scattered pulsars, which generally have much more dynamic lines of sight, are feasible for pulsar timing arrays. 
\par This resolution, combined with the tens-to-hundreds of lines of sight observed at sub-monthly cadences, means pulsar timing arrays can, as ancillary science, perform simultaneous studies of a significant fraction of the local galaxy with high temporal resolution and years-long baselines without sacrificing the timing precision necessary for their primary science. Relatedly, the ability to now resolve a significant range of scintillation bandwidths means that scattering delays can now be estimated for almost all pulsars, even if direct measurements via cyclically-deconvolved pulse broadening functions are not feasible, which is invaluable for informing pulsar timing array noise models for the purpose of gravitational wave sensitivity. Rather than being limited to estimations for pulsars with delays of typically less than 50 ns \citep{Levin_Scat,turner_scat,epta1}, which can be considered inconsequential at current pulsar timing array sensitivities \citep{Agazie_2023_noise}, we can now monitor significantly larger scattering delays that may actually have a noticeable impact on pulsar timing array noise budgets.
\par For these reasons, should observing constraints arise, it is crucial for pulsar timing arrays to strategize which pulsars they want to prioritize for cyclic spectroscopy. While pulsars such as PSR B1937+21 have demonstrated success with this technique, both from a cyclic deconvolution standpoint \citep{cyc_spec,wdv13} as well as with scintillation bandwidth measurements and scattering delay estimations \citep{Turner_1937}, its red noise overwhelmingly dominates its timing residuals despite also being fairly highly scattered \citep{Agazie_2023_noise}. Conversely, pulsars like J1643--1224, which exhibits residuals on the order of microseconds \citep{Agazie_2023_timing}, appears to have its timing noise dominated by scattering, primarily due to its positioning behind an HII region relative to Earth \citep{Harvey-Smith_2011, Ocker_2020,mall,ocker}. As mentioned earlier, scintillation bandwidth-estimated scattering delays for this pulsar are at comparable levels to these timing residuals. 
\begin{figure*}[!ht]
    \centering
    \captionsetup[subfigure]{labelformat=empty}
    \subfloat[]{\hspace*{-.8cm} {\includegraphics[width=0.55\textwidth]{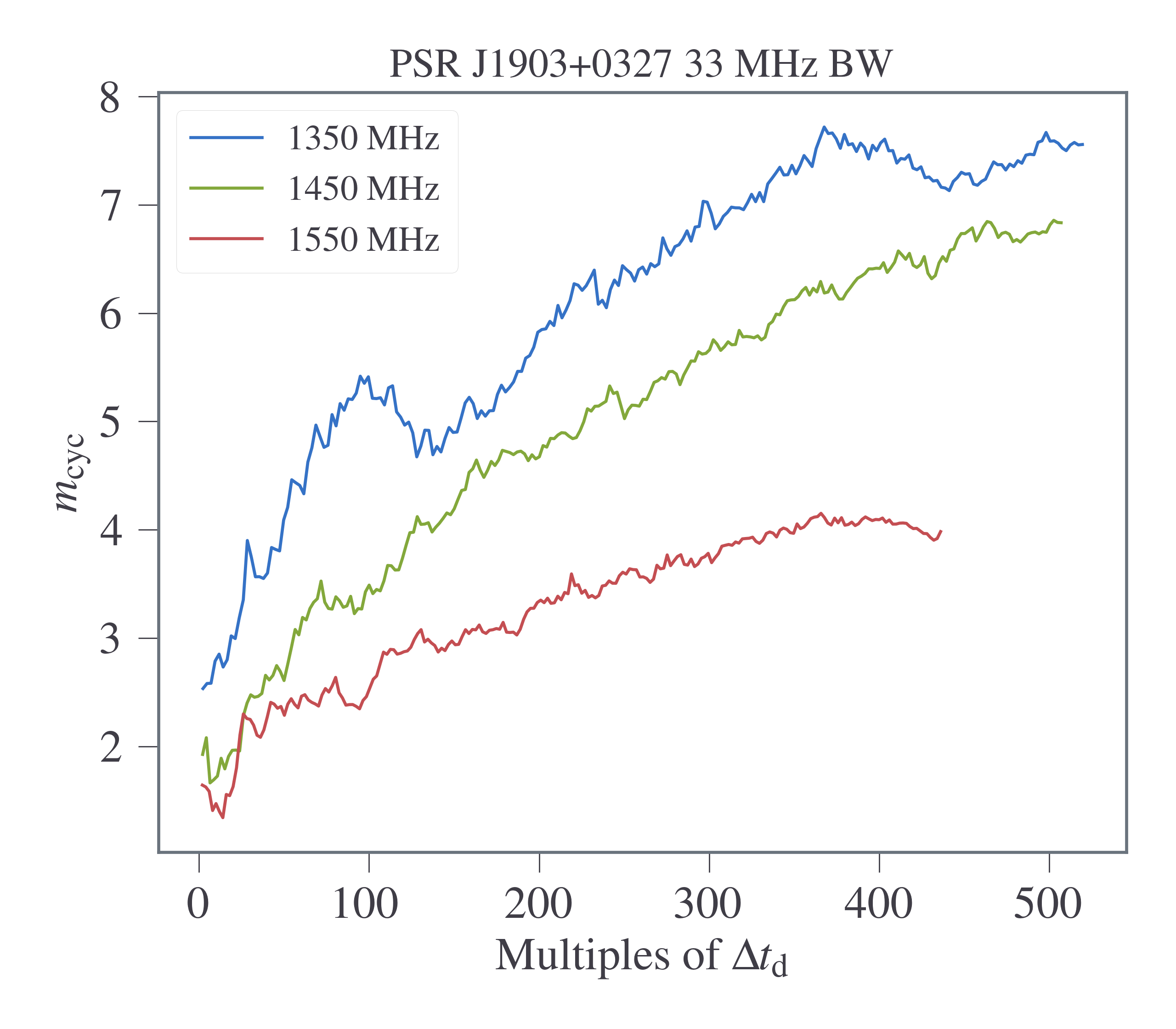} }}%
    {\hspace*{-.4cm}\includegraphics[width=0.55\textwidth]{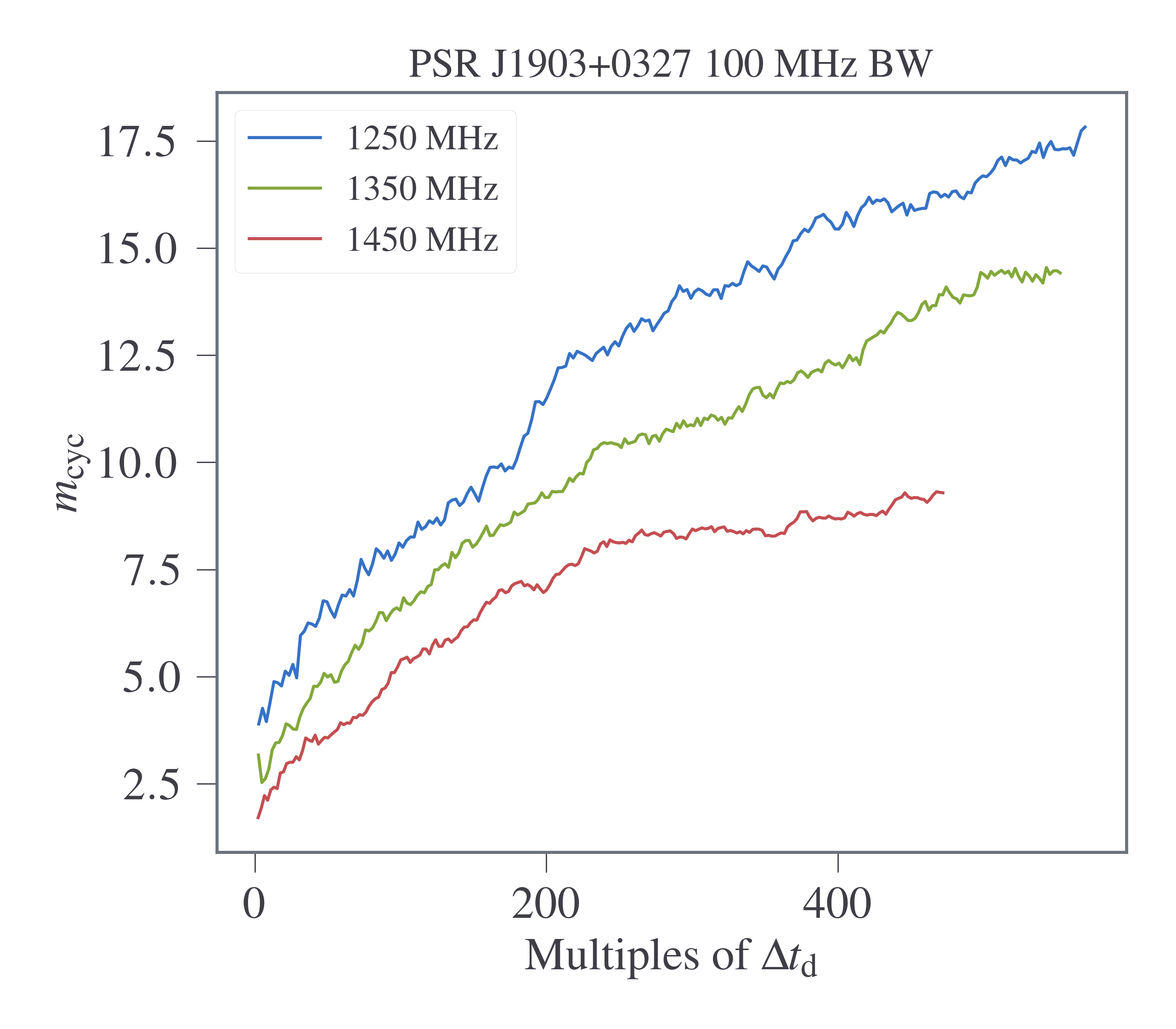} }%
    \caption{Cyclic merit over multiples of the scintillation timescale at different observing frequencies with 33 (left) and 100 (right) MHz of bandwidth for PSR J1903+0327. The low cyclic merit even after long integration times suggests significantly more sensitive instruments will be required for this pulsar to benefit from cyclic spectroscopy. Note that these cyclic merits were calculated with observational data, and so cumulative S/N is not guaranteed to constantly increase with additional subintegations.}%
    \label{many_scint_time}%
\end{figure*}
\par Finally, pulsars that initially seem like prime candidates for benefiting from cyclic spectroscopy may in actuality not be accessible to the technique with most currently available instrumentation. As shown in Table \ref{regime_table}, PSR J1903+0327 is well within the full deconvolution regime at L--band, where pulsar timing arrays primarily operate. Additionally, its scattering delays have been directly measured at 10s$-$100s of microseconds in this frequency range \citep{geiger2024nanograv125yeardataset} while simultaneously exhibiting microsecond scale pulse time-of-arrival uncertainties and residuals \citep{Agazie_2023_noise}. However, we were unable to achieve any cyclic deconvolution, partial or otherwise, in the three 2-hr observations we took of this pulsar, despite estimating a modest cyclic merit that was greater -- by over an order of magnitude in some cases -- than pulsars for which we saw partial phase retrieval. This is in large part due to this pulsar's cyclic merit being dominated by its large scattering delay, which significantly makes up for its low S/N. While achieving a S/N of 16 may be possible over integration times of tens of minutes, it will be challenging with current instrumentation to achieve this sensitivity on timescales required for cyclic deconvolution.
\par This can be further understood by examining this pulsar's estimated scintillation bandwidth in this frequency range, which is around 0.86 kHz, as shown in Table \ref{regime_table}. Assuming a minimum of three frequency channels to claim a detection \citep{turner_scat}, we would require 285 Hz resolution at L--band, necessitating around 5480 cyclic channels per polyphase filterbank channel assuming we maintain NANOGrav's standard setup. 
\par While this configuration does not constitute an excessive number of cyclic channels from a computational standpoint, it does from a S/N standpoint; PSR J1903+0327 is already a fairly dim pulsar to begin with, recording flux densities of around 0.7 mJy at L--band \citep{alam2020nanograv}. For context, the next dimmest pulsar we observed, PSR J1643--1224, is over seven times brighter. To achieve the channelization needed to attempt scintillation measurements for PSR J1903+0327, per the radiometer equation, we would need to sacrifice around three times as much signal per scintle from the frequency resolution with which we analyzed the other two pulsars we observed. However, even at that coarser channelization, we already were unable to observe any signal in PSR J1903+0327. In fact, for a fully integrated 90-minute observation, while keeping the same number of pulse phase bins, we were only able to see the pulsar in the intensity profile at very low S/N after we frequency averaged to a resolution of 25 kHz, which has around 95 times more signal per scintle than our required cyclic configuration. For effective cyclic deconvolution, or even to simply construct a dynamic wavefield, we need to limit each recovered transfer function to integration times of less than a pulsar's scintillation timescale, which, for PSR J1903+0327, is expected to be less than 15 seconds, per estimations by the NE2001 electron density model \citep{NE2001}. Modifying our observing setup to the required observing frequency resolution and decreasing our subintegration times from 25 seconds to 15 seconds would result in around 122 times less signal per scintle than the aforementioned configuration in which PSR J1903+0327 was barely visible.

\par If we attempted to compensate by observing in NANOGrav's next lowest frequency band to take advantage of a greater observed flux density, we would observe a smaller scintillation bandwidth and even less S/N per scintle ($\propto 1/\sqrt{\rm{Chan. \ BW}}$). Conversely, if we attempted to observe in NANOGrav's next highest frequency band, while we would have a larger scintillation bandwidth, the observed flux density would be lower by more than 30\% on average. Naively, one could attempt to circumvent the low S/N issue by simply ignoring the scintillation timescale limitation in an attempt to achieve greater S/N. However, this would require significantly longer observing times than the $\sim$30 minutes that PTAs usually reserve for observations. As shown in Figure \ref{many_scint_time}, with the GBT, integrations over hundreds of scintillation timescales is insufficient to even break a cyclic merit of 10 with our chosen observing bandwidth, although increasing this bandwidth to 100 MHz does allow for sufficiently large cyclic merits in these same periods of time. That being said, given how highly scattered this pulsar is, evolution in scintle size across such a large bandwidth may be significant enough to render this option invalid, even before considering whether a transfer function could be properly recovered after integrating over so many scintillation timescales. These cyclic merits also do not account for the low S/N per scintle that we would be limited to at the channelizations necessary to resolve scintillation.
\par For these reasons, unless cyclic spectroscopy is performed on observations from instruments significantly more sensitive than the Green Bank Telescope, this pulsar currently has far too low of a S/N for successful resolution of scintles, much less cyclic deconvolution. This conclusion is also in line with the findings of \cite{dsj+20}, as they too estimated a low cyclic merit. This lack of S/N could potentially be remedied with more sensitive telescopes, including current instruments like FAST, or upcoming instruments like DSA-2000. 
\par To mitigate these discrepancies between modest cyclic merit and lack of results, we introduce a cyclic merit 2.0, $m_{\rm cyc, 2.0}$. This metric behaves identical to the original cyclic merit and follows the same trends seen throughout this paper, but relies on scintillation bandwidth rather than observing bandwidth. Additionally, with S/N now relying on flux density over a single scintle rather than the width of the observing band, our new metric will now prefer pulsars that exhibit visible scintles, rather than pulsars that simply exhibit large scattering delays. 

\begin{figure*}[!htp]
    \centering
    \captionsetup[subfigure]{labelformat=empty}
    {\includegraphics[width=1.0\textwidth]{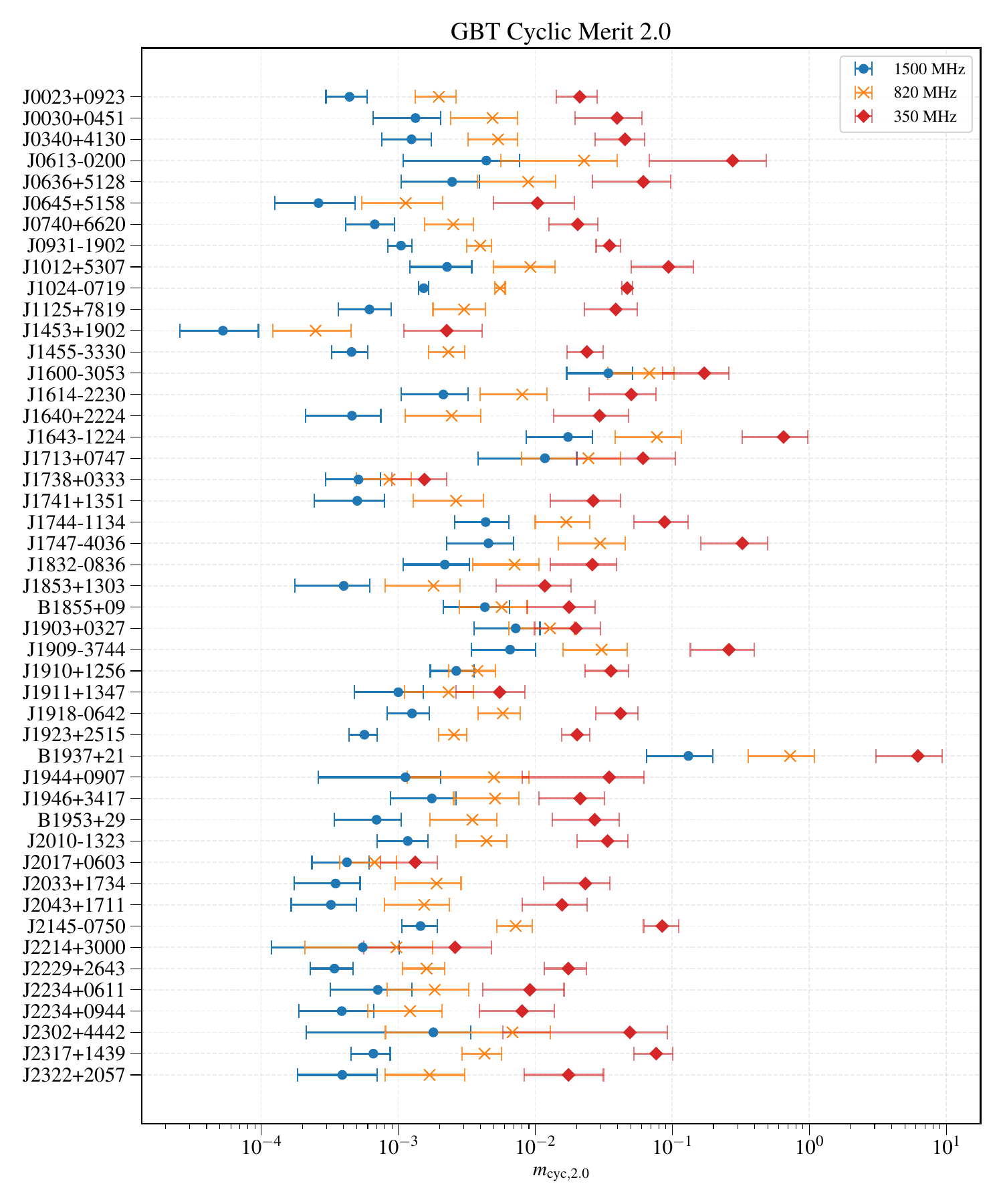} }%
    \caption{Estimated cyclic merit 2.0 for pulsars in the NANOGrav 12.5-year data set at observing frequencies of 350 (red diamonds), 820 (orange crosses), and 1500 (blue circles) MHz assuming data were taken with the GBT. We would expect full cyclic deconvolution for sources with $m_{\rm cyc, 2.0}\gg 0.05$.}%
    \label{cyc_2_gbt}%
\end{figure*}

\begin{figure*}[!htp]
    \centering
    \captionsetup[subfigure]{labelformat=empty}
    {\includegraphics[width=1.0\textwidth]{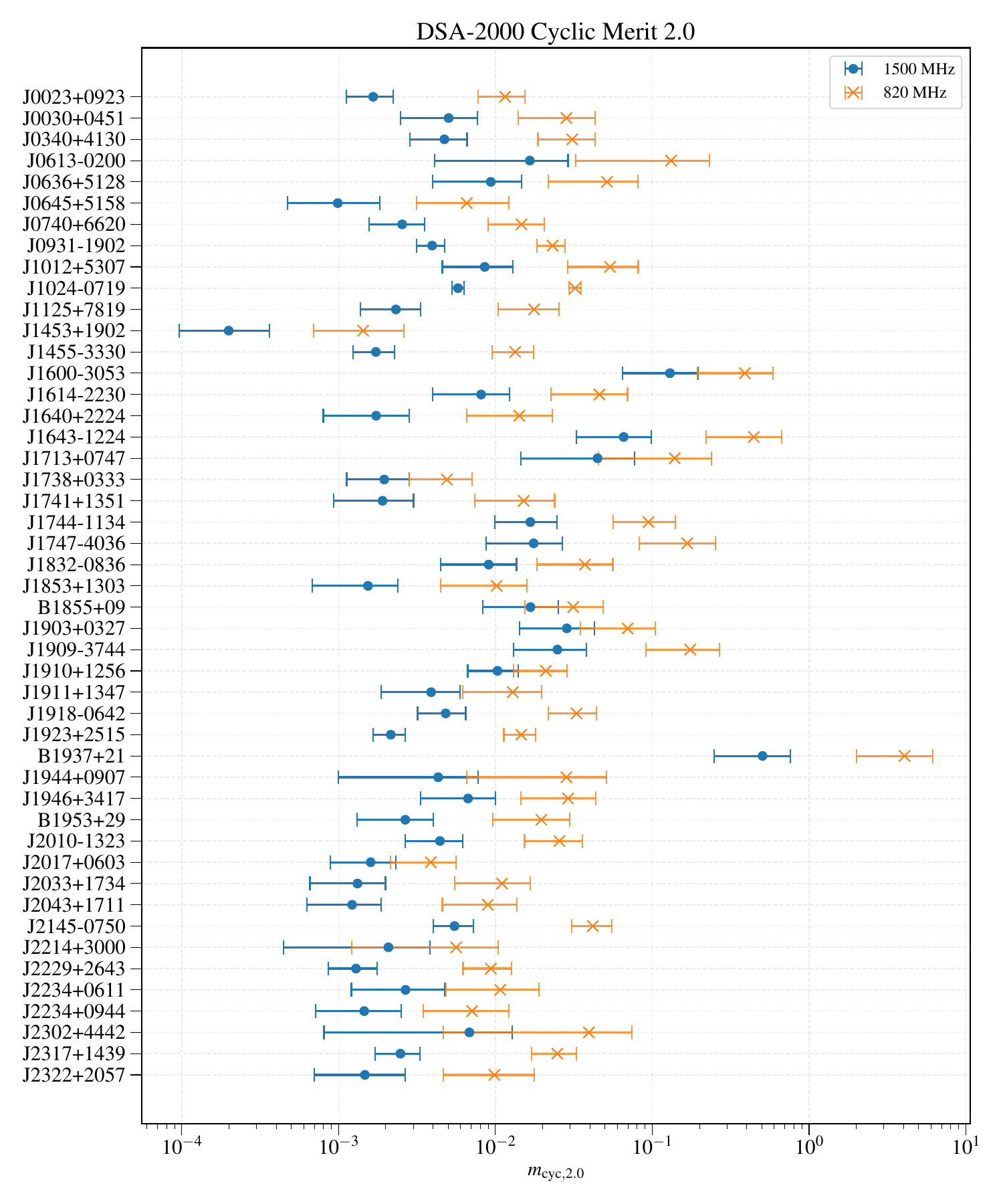} }%
    \caption{Estimated cyclic merits 2.0 for pulsars in the NANOGrav 12.5-year data set at observing frequencies of 820 (orange crosses) and 1500 (blue circles) MHz assuming data were taken with DSA-2000. We would expect full cyclic deconvolution for sources with $m_{\rm cyc, 2.0}\gg 0.05$.}%
    \label{cyc_2_dsa}%
\end{figure*}
\par To examine how this new metric affects calculated cyclic merit values, we redid our cyclic merit calculations for all NANOGrav pulsars analyzed in \ref{nano_cyc}. The results for the GBT and DSA-2000 can be seen in Figures \ref{cyc_2_gbt} and \ref{cyc_2_dsa}, respectively. Assuringly, under this new metric, PSR B1937+21 still has far and away the highest cyclic merit, with PSR J1643$-$1224 still in second place, while PSR J1903+0327 is effectively lost amongst the majority of NANOGrav pulsars. Interestingly, PSR J1909$-$3744 now holds one of the highest cyclic merits at 350 MHz, likely in large part due to a fairly large scintillation bandwidth resulting in significant S/N across a given scintle, alongside its comparatively small pulse width and effective pulse width. This result is further justified given that 350 MHz is fairly close to this pulsar's expected full deconvolution regime threshold, which, according to Figure \ref{decon_plot}, should be around 250 MHz.

\par
\begin{figure}[!ht]
    \centering
    \hspace*{-1cm}                                                           
    \includegraphics[scale = 0.6]{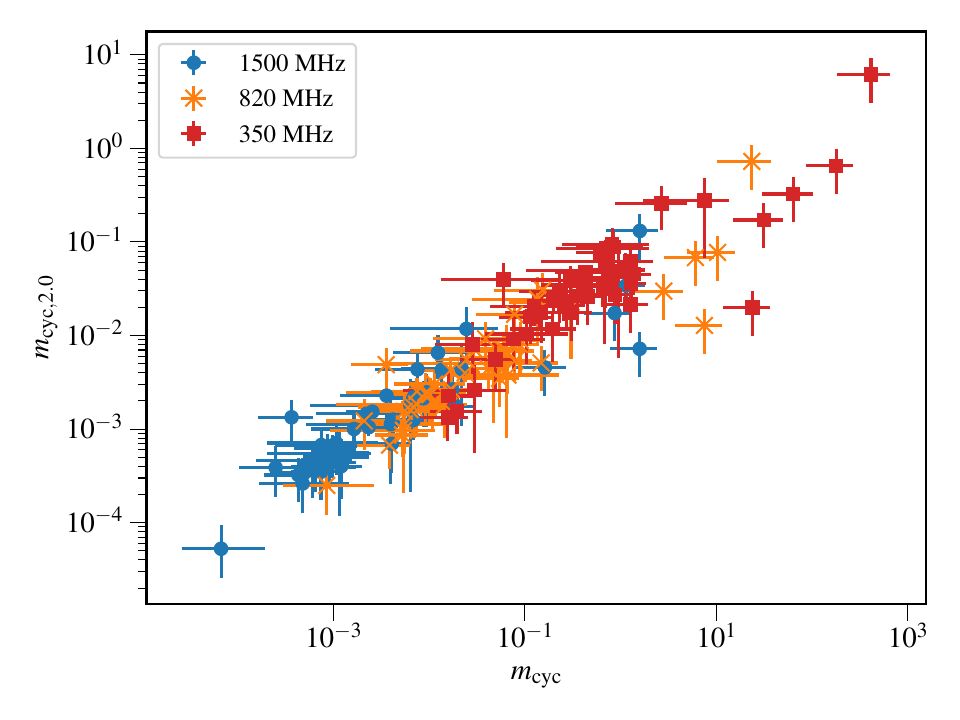}
    \caption{Cyclic merit 1.0 (x-axis) and 2.0 (y-axis) at 350 (red squares), 820 (orange crosses), and 1500 (blue circles) MHz for all pulsars in NANOGrav's 12.5-year data set. A simple power law fit indicates that the original $m_{\rm cyc}\gg 1$ threshold corresponds approximately to $m_{\rm cyc, 2.0} \gg 0.05$.}
    \label{cyc_compare}
\end{figure}

 To determine how cyclic merit 2.0 can be used to estimate the likelihood of successful cyclic deconvolution, we compared how a given pulsar's calculated cyclic merit 1.0 differed from its corresponding cyclic merit 2.0 at a given observing frequency, with the results shown in Figure \ref{cyc_compare}. By examining this behavior in log space, we can see that the resulting trend can be described quite well by a simple power law fit. When performing such an analysis in aggregate, our fit indicates that the original cyclic deconvolution threshold  of $m_{\rm cyc}\gg 1$ translates to $m_{\rm cyc, 2.0} \gg 0.05$.

\break
\section{Conclusions \& Future Work}
\label{concl}
We explore the advantages and impediments to analyzing cyclic spectroscopy-processed data in regimes where full deconvolution of the transfer function from the intrinsic pulse is not possible. Using data in both full- and partial-deconvolution regimes, we obtain estimates of the ``boundary" between these  regimes, as well as cyclic merit. We show that the majority of NANOGrav pulsars observed in the 12.5-year data release are observed by the PTA in the partial-deconvolution regime, and the five pulsars for which full deconvolution appears to be possible above 300 MHz would likely need to be observed at lower frequency ranges than are utilized in NANOGrav's current observing strategy with the GBT. That being said, significant benefits are still obtainable with cyclic spectroscopy by NANOGrav continuing  observations in the partial-deconvolution regime. 
\par Through examinations of the evolution of cyclic merit and recovered phase over observing frequency, we show that cyclic deconvolution is a continuous process, and that more information can be recovered the closer one observes to the full deconvolution regime. We also construct dynamic and secondary wavefields, as well as dynamic and secondary spectra, and discuss what can be done with partial phase retrieval. We find that the addition of recovered phase, even if incomplete, provides valuable additional signal information that can be used for scattering delay estimation as well as studies of the interstellar medium. We conclude that the simultaneous high time and frequency resolution alone is sufficient to justify preferring this processing technique over Fourier spectroscopy, even in instances where S/N is too low to attempt phase retrieval. Finally, we conclude that, despite being observed in the full-deconvolution regime, highly scattered but low flux density sources like PSR J1903+0327, will not be able to benefit from cyclic spectroscopy unless observed with significantly more sensitive instruments than the GBT. We then use this as motivation to introduce a new cyclic merit 2.0, which considers cyclic merit over individual scintles rather than the observing band.
\par As of this writing, efforts are almost complete on a cyclic spectroscopy backend at the Green Bank Observatory. This will mark the first time that cyclic spectroscopy-processed data will be available for general use without the need for long term storage of baseband data or knowledge of current processing software, making the benefits of the technique significantly more accessible. Even though pulsar timing arrays primarily operate outside of the full-deconvolution regime, once they adopt this technique, sensitivity to gravitational waves should improve noticeably in highly scattered pulsars thanks to the ability to actually estimate their scattering delays. Additionally, expansive studies of the interstellar medium will now be possible thanks to the significant improvements in frequency resolution, many views through the interstellar medium, and years-long timing baselines available through these larger scale research efforts. 
\section{Acknowledgments} 
\par We gratefully acknowledge support of this effort from the NSF Physics Frontiers Center grants 1430284 and 2020265 to NANOGrav. Some of the data processing in this work utilized the resources of the Link computing cluster at West Virginia University. The National Radio Astronomy Observatory and Green Bank Observatory are facilities of the U.S. National Science Foundation operated under cooperative agreement by Associated Universities, Inc. We thank the AO staff for their assistance during the observations reported in this work, in particular P. Perillat, J. S. Deneva, H. Hernandez, and the telescope operators. TD and DRS were partially supported through the National Science Foundation (NSF) PIRE program award number 0968296. TD acknowledges NSF AAG award number 2009468, sabbatical and summer leave funding from Hillsdale College, and the Hillsdale College LAUREATES program. We thank Ross Jennings for valuable discussion and Fengqiu Adam Dong for paper comments.

\par \textit{Software}: \textsc{dspsr} \citep{dspsr}, \textsc{astropy} \citep{astropy}, \textsc{pypulse} \citep{pypulse}, \textsc{scintools} \citep{scintools}, \textsc{PyGDSM} \citep{2016ascl.soft03013P}, \textsc{numpy} \citep{numpy}, and \textsc{matplotlib} \citep{matplotlib}.
\bibliography{turner_1937_cs.bib}{}
\bibliographystyle{aasjournal}
\end{document}